\documentclass[acmtog,nonacm]{acmart}
\setcopyright{none}
\authorsaddresses{}

\usepackage{booktabs}
\usepackage[ruled]{algorithm2e} 

\SetAlFnt{\small}
\SetAlCapFnt{\small}
\SetAlCapNameFnt{\small}
\SetAlCapHSkip{0pt}

\usepackage[utf8]{inputenc} % allow utf-8 input
\usepackage[T1]{fontenc}    % use 8-bit T1 fonts
\usepackage{hyperref}       % hyperlinks
\usepackage{url}            % simple URL typesetting
\usepackage{booktabs}       % professional-quality tables

\usepackage{amsfonts}       % blackboard math symbols
\usepackage{amssymb}
\usepackage{amsmath}
\usepackage{amsthm}
\usepackage{bbm}

\usepackage{nicefrac}       % compact symbols for 1/2, etc.
\usepackage{microtype}      % microtypography
\usepackage{lipsum}		% Can be removed after putting your text content
\usepackage{graphicx}
\usepackage{doi}

\usepackage{xcolor}
\usepackage{mdframed}
\usepackage{hhline}

\usepackage{graphicx}

\usepackage[font=small,labelfont=bf]{caption}
\definecolor{c1}{gray}{0.96}
\definecolor{c2}{gray}{0.70}
\definecolor{c3}{gray}{0.40}
\definecolor{c4}{gray}{0.48}
\definecolor{hlcolor}{rgb}{0.55,0.2,0.3}
\definecolor{emphcolor}{rgb}{0.4,0.6,0.6}

\hypersetup{
    colorlinks=true,       % false: boxed links; true: colored links
    linkcolor=red,          % color of internal links (change box color with linkbordercolor)
    citecolor=green,        % color of links to bibliography
    filecolor=magenta,      % color of file links
    urlcolor=emphcolor      % color of external links
}

\newcommand{\txtbold}[1]{\textbf{{\sffamily
\color{c3}{#1}}}}

\newcommand{\figcaptionwide}[3]{
\begin{figure*}[ht]
  \includegraphics[width=\textwidth]{#1}
  \caption{#2}
  \label{#3}
\end{figure*}
}
\newcommand{\figcaption}[3]{
\begin{figure}[ht]
  \centering
  \includegraphics[width=\linewidth]{#1}
  \caption{#2}
  \label{#3}
\end{figure}
}

\theoremstyle{definition}

\mdfdefinestyle{mdlight}{
  hidealllines=true,backgroundcolor=c1,
  innerleftmargin=8pt,innerrightmargin=8pt,innertopmargin=8pt,
  rightmargin=10pt,skipabove=0,skipbelow=0}
\mdfdefinestyle{mddark}{
  hidealllines=true,backgroundcolor=c2,
  innerleftmargin=15pt,innerrightmargin=15pt,innertopmargin=8pt,
  rightmargin=5pt,}
\surroundwithmdframed[style=mdlight]{ppr}
\surroundwithmdframed[style=mdlight]{dsc}

\newcommand{\pref}[3]{\begin{mdframed}[style=mdlight]
\txtbold{\href{#2}{[x]} #1}}

\usepackage{enumitem}
\setlist[itemize]{noitemsep,nosep}
\begin{document}
\title{Democratizing the Creation of Animatable Facial Avatars}

\author{Yilin Zhu}
\affiliation{
 \institution{Stanford University}
 \country{USA}}

\author{Dalton Omens}
\affiliation{
 \institution{Stanford University}
 \country{USA}}
\affiliation{
 \institution{Epic Games}
 \country{USA}}

\author{Haodi He}
\affiliation{
 \institution{Stanford University}
 \country{USA}}

\author{Ron Fedkiw}
\affiliation{
 \institution{Stanford University}
 \country{USA}}
\affiliation{
 \institution{Epic Games}
 \country{USA}}

\renewcommand\shortauthors{Zhu, Y. et al}

\begin{abstract}
In high-end visual effects pipelines, a customized (and expensive) light stage system is (typically) used to scan an actor in order to acquire both geometry and texture for various expressions. Aiming towards democratization, we propose a novel pipeline for obtaining geometry and texture as well as enough expression information to build a customized person-specific animation rig without using a light stage or any other high-end hardware (or manual cleanup). A key novel idea consists of warping real-world images to align with the geometry of a template avatar and subsequently projecting the warped image into the template avatar's texture; importantly, this allows us to leverage baked-in real-world lighting/texture information in order to create surrogate facial features (and bridge the domain gap) for the sake of geometry reconstruction. Not only can our method be used to obtain a neutral expression geometry and de-lit texture, but it can also be used to improve avatars after they have been imported into an animation system (noting that such imports tend to be lossy, while also hallucinating various features). Since a default animation rig will contain template expressions that do not correctly correspond to those of a particular individual, we use a Simon Says approach to capture various expressions and build a person-specific animation rig (that moves like they do). Our aforementioned warping/projection method has high enough efficacy to reconstruct geometry corresponding to each expressions.
\end{abstract}

\maketitle

%%%%%%%%% INTRO
\section{Introduction}
\label{sec:intro}

The use of personalized avatars has become increasingly prevalent in a wide range of applications: massively multiplayer online games (e.g. Roblox), the entertainment and animation industry (e.g. MetaHumans from Epic Games), virtual/augmented reality experiences (e.g. Apple, Meta, Snapchat), and video conferencing platforms (e.g. Microsoft Teams, Zoom). 
Personalized avatars have been shown \cite{sailer2017gamification} to enhance user engagement and satisfaction in a variety of online settings including gaming, virtual reality, and metaverse applications.
As the concept of a metaverse gains traction, avatars are poised to become even more ubiquitous; however, the public still faces barriers to creating personalized avatars, such as limited access to high-end capture hardware and/or the artist tools/expertise required to hand-craft such avatars. 

With democratization in mind, we focus on the utilization of technologies that are currently widespread and are expected to be ubiquitous going forward, such as cell phones and webcams. Our pipeline intentionally avoids the use of detailed depth data and scanning, which has slowly been gaining popularity (e.g.\ because of the Kinect) but also at the same time losing ground in important hardware devices (e.g. the quality of depth data from the iPhone front-facing TrueDepth camera has been decreased in recent versions, and will likely be further decreased minimizing its use to security for the phone unlock feature). On the other hand, our pipeline embraces the continually improving RGB cameras in cell phones and webcams. We mostly rely on tasks executable by a non-expert user, e.g.\ asking for images from approximate angles, a turn-table video made while maintaining a neutral expression, and images of expressions that are prompted via a visual Simon Says approach.

Although there are a variety of recent and interesting approaches to representing facial avatars (e.g.\ using NeRFs \cite{mildenhall2021nerf}, PIFu \cite{saito2019pifu}, etc.\! ), our pipeline reconstructs explicit geometry and de-lit texture so that it can be utilized in the widest number of existing graphics applications. The high-end special effects companies (e.g.\ ILM, Weta, Digital Domain, Pixar) typically use proprietary in-house software; however, there have been various efforts to democratize facial animation. Maya \cite{maya} has various plugins that make facial animation easier, but one still has to sculpt/provide the face geometry/texture (and joints need to be placed by hand). Rigify \cite{rigify} in Blender \cite{blender} similarly, requires that one provide geometry/texture and place all the joints by hand. There are two standalone softwares, Headshot2 from Reallusion \cite{Headshot2} and FaceGen from Daz3D \cite{Daz3D}, that create animatable geometry/texture without requiring the user to provide it, but the quality is hindered by a reconstruction that only uses a single front-facing image (no side or three-quarters images). MetaHumans \cite{mhc} in the Unreal Engine \cite{unrealengine} is the only publicly available tool that enables the creation of animatable geometry using more than just a single front-facing image, i.e.\ Mesh2Metahuman \cite{m2m} creates an animatable MetaHuman from input 3D geometry; unfortunately, there is no algorithm for obtaining texture (and one has to hand select from a library of defaults/sliders).

Although there is a plethora of prior work that creates floating face masks (often without texture, making it useless for preserving likeness), this geometry needs to be imported into animation software for subsequent use. Missing features (e.g.\ eyes, ears, inner lip and mouth, hair, etc.\!  ) need to be added manually or hallucinated, and the geometry needs to be retopologized (which can be lossy). Since this hallucination/lossiness modifies the likeness, it is essential that one be able to improve upon the likeness of the facial model after any such import. Methods that infer geometry/texture from photos cannot do this unless they are trained to infer animatable models directly (see e.g.\ \cite{shi2020fast}\cite{lin2021meingame}). Methods that solve an optimization problem to determine vertex positions (or parameterizations, e.g.\ morphable models  \cite{blanz1999morphable}, FLAME\cite{li2017learning}, etc.\! ) and texel colors can typically be modified to improve the likeness of the facial model after the import.

It is well-known that geometry/texture reconstruction is best evaluated from novel views (see figure \ref{fig:good_tex_bad_geo}), but less discussion occurs around the fact that animated expressions further highlight inaccuracies in geometry reconstruction and texture alignment (see figure \ref{fig:bad_geo}). A more subtle but also problematic issue is the mismatch between the template expressions on the animation rig and the corresponding expressions of an individual. The way someone smiles, frowns, speaks, etc.\ is part of their motion signature and intrinsic to their likeness. We address this by using a Simon Says approach to prompt the user into producing various expressions that are reconstructed with our warping/projection approach; subsequently, the results are used to create a person-specific animation rig with motion signatures that better preserve the likeness of an individual.

\figcaption{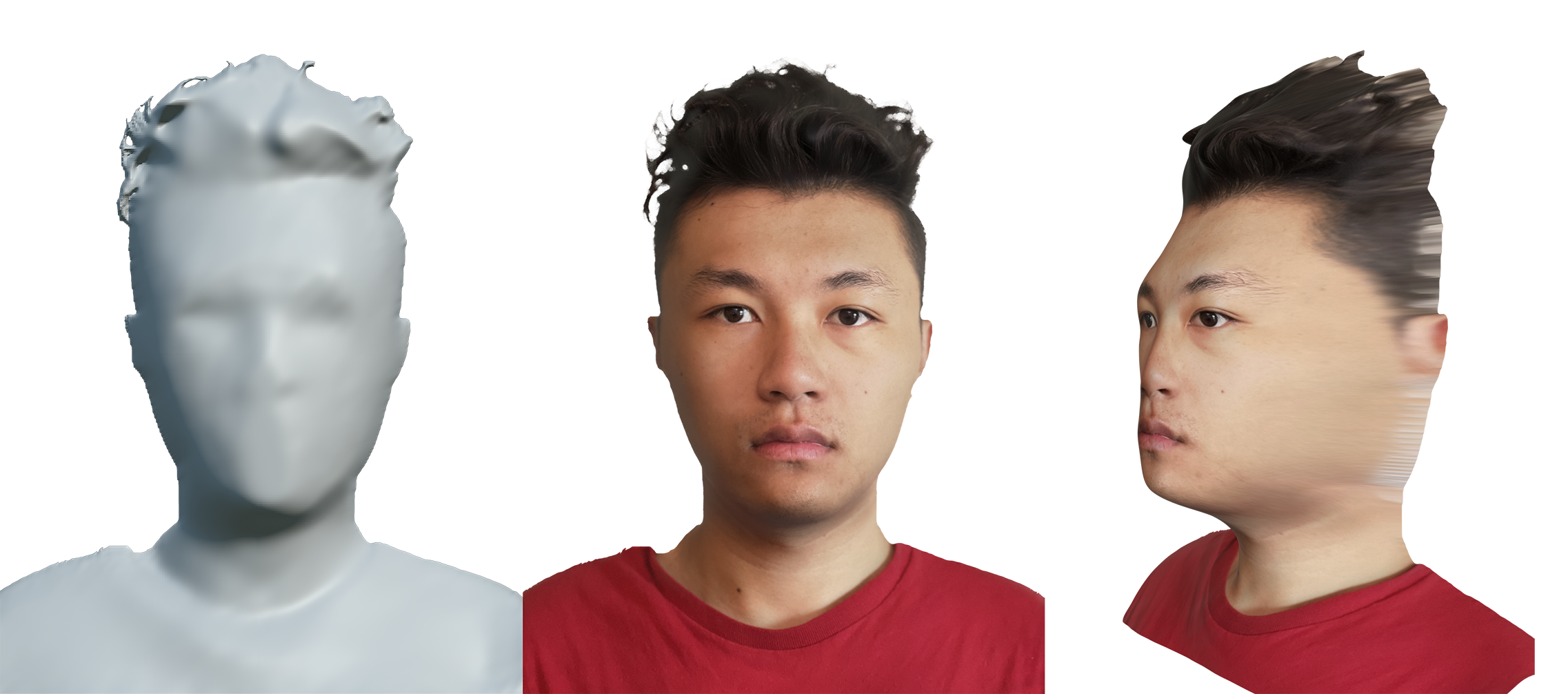}{From left to right: geometry, same geometry textured from a front-facing image, same geometry/texture as seen from a three-quarters view. This emphasizes how misleading a textured geometry can be when not considered from significantly novel views.}{fig:good_tex_bad_geo}

\figcaption{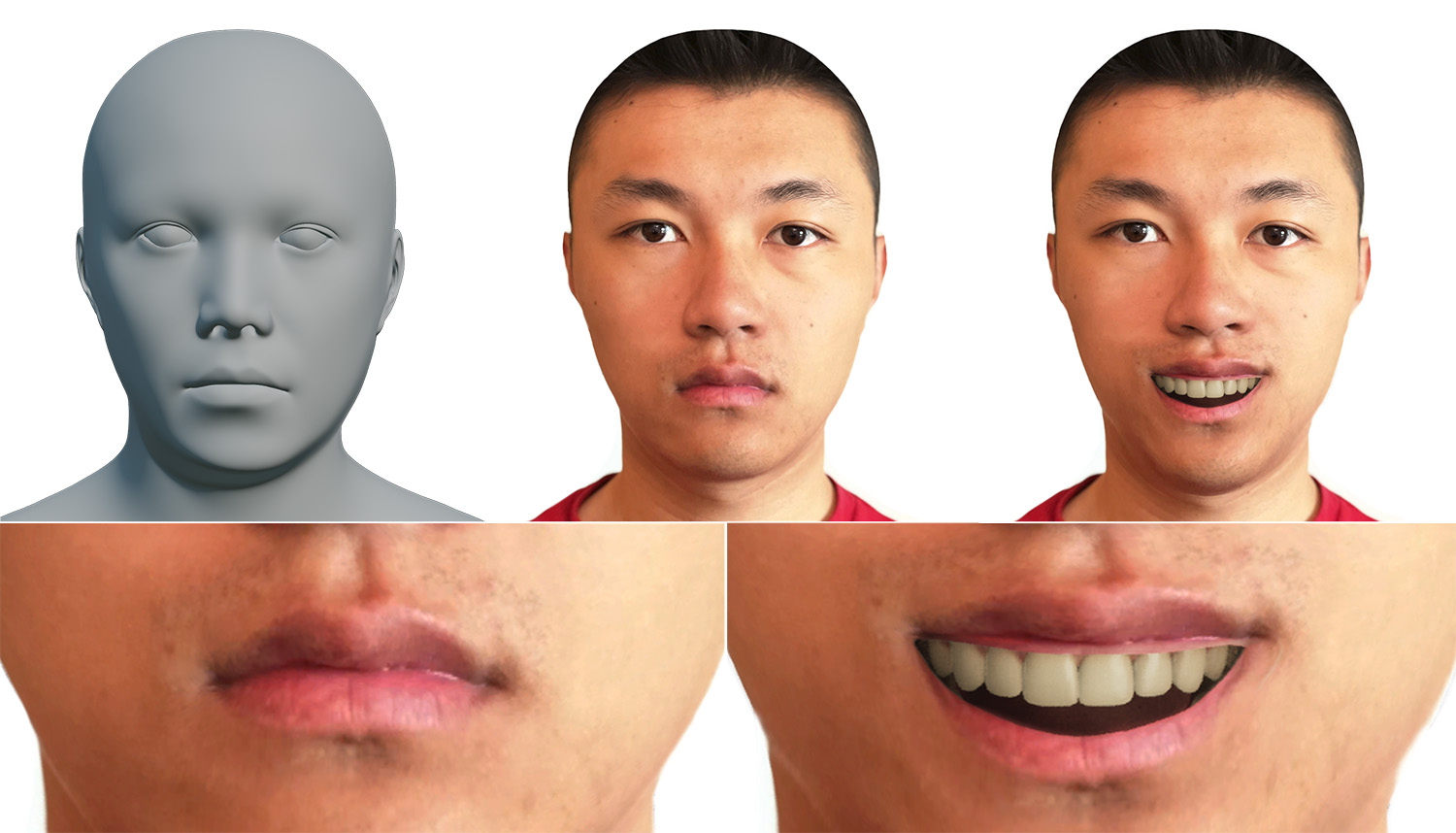}{From top left to top right: avatar geometry (derived from the geometry in Figure \ref{fig:good_tex_bad_geo}), same geometry textured from a front-facing image, same geometry/texture with a smile expression. This emphasizes how misleading a textured geometry can be when not considering various expressions. Bottom: zoomed-in view of the top middle and top right figures. Note in particular how part of the bottom lip texture (and the crease between the lips) appears on the top lip.}{fig:bad_geo}

The novel contributions of our pipeline can be summarized as follows:
\begin{itemize}
    \item We intentionally use baked-in lighting to create surrogate features by warping and projecting real-world images into the synthetic geometry's texture before subsequent optimization (of the geometry); notably, each view gets its own surrogate features (i.e.\ texture).
    \item We import our reconstructed geometry into the MetaHuman animation system, noting that this import is both lossy (i.e.\ modifies the input geometry) and hallucinates various features (e.g.\ eyes, ears, inner lip and mouth, hair, etc.\! ). Notably, this is the only publicly available option (we are aware of) that non-experts have for creating an animatable avatar from an input mesh. Subsequently, we show that our warping/projection approach can be used to improve the likeness of this imported animatable geometry. 
    \item We propose a method for creating well-aligned de-lit textures from warped and projected images, preserving details such as moles, freckles, and stubble (important for maintaining likeness) while removing baked-in lighting. Importantly, our approach can be used to provide textures for animatable MetaHumans (which currently only utilize a library of defaults/sliders).
    \item We morph a default animation rig (similar to \cite{lin2022leveraging}) to match the neutral expression geometry, and subsequently use a visual Simon Says approach to prompt the user into producing various expressions that are reconstructed (via our warping/projection approach) and used to build a person-specific animation rig (preserving motion signature likeness).
\end{itemize}

Section \ref{sec:initialgeo} discusses how we obtain an initial reconstruction of the geometry in order to bootstrap our pipeline. Section \ref{sec:neutralgeo} and Section \ref{sec:neutraltex} discuss our novel techniques for creating high-efficacy neutral/expressionless geometry and high resolution de-lit textures respectively. Notably, the result of Section \ref{sec:neutralgeo} and Section \ref{sec:neutraltex} is a fully animatable rig (with eyes, mouth, hair, etc.\! ), not just a floating face mask. Section \ref{sec:video} addresses the use of our method when one does not have access to the subject that one wants to reconstruct. Section \ref{sec:rigmorph} discusses the fitting of a template animation rig to the reconstructed geometry, as well as how our Simon Says framework can be used to create a person-specific animation rig that better preserves an individual's motion signatures. 

%%%%%%%%% RELATED WORK
\section{Related Work}
\label{sec:related}

In addition to the works discussed below, we refer the reader to the following survey papers: \cite{chrysos2018comprehensive} \cite{zollhofer2018state} \cite{tewari2022advances} \cite{tretschk2023state}.

\textbf{High-End Light Stages:}
The light stage \cite{debevec2000acquiring} \cite{debevec2012light} \cite{liu2022rapid} uses hundreds of light/camera combinations to acquire a 4D reflectance field, enabling the highest quality facial reconstruction currently available.
The light stage and similar systems \cite{ghosh2011multiview} \cite{joo2017panoptic} \cite{hendler2018avengers} \cite{zhang2022video} \cite{bolkart2023instant} that directly control illumination capture extremely high-fidelity geometry and separate albedo/specular/displacement maps. 
In addition to capturing a neutral/expressionless geometry and texture, these systems can be used to capture various facial expressions; subsequently, a visual effects artist (after some manual cleanup) can construct a high-quality facial animation rig.  Moving towards democratization, various efforts have focused on reducing the required number of lights and/or cameras while maintaining high-quality results (e.g.\ \cite{zhang-siggraph2004-stfaces}); in particular, \cite{lattas2022practical} uses (easily portable) commodity components to construct facial capture systems that sit on a typical office desk.

\textbf{High-End Multi-view:}
Various high-end systems \cite{medusa} \cite{beeler2010high} \cite{bradley2010high} \cite{riviere2020} lack the high degree of lighting control available to a light stage, but still use a number of carefully calibrated cameras (intrinsics and extrinsics) along with measured/controlled lighting in a laboratory setting. Such systems are often preferable (to light stages) for performance capture, since a light stage can feel constraining to an actor.

\textbf{Single-view Democratization:}
The typical scenario for democratization is one in which the user has access to a single camera or cellphone (perhaps with a tripod) and no control over the lighting (but perhaps access to a cheap version of a chrome sphere, such as an ornament); interestingly, it is possible to program an at-home large-screen TV in order to mimic various light patterns from a light stage (see also \cite{sengupta2021light}). In this democratized setting, noise and other inaccuracies (imperfect camera calibration, unknown lighting, etc.\! ) can lead to disappointing results; thus, a strong prior on allowable face shapes is typically required. Although the 3D Morphable Model (3DMM) \cite{blanz1999morphable} has historically been the most prevalent method of choice, recent works (such as FLAME \cite{li2017learning} and  \cite{booth20163d} \cite{tewari2017mofa} \cite{tran2018nonlinear} \cite{wang2022faceverse} \cite{chandran2022facial}) have aimed to improve upon 3DMMs. Either optimization \cite{romdhani2003efficient} \cite{li2017learning} or deep learning \cite{richardson20163d} \cite{zhu2016face} \cite{bulat2017far} \cite{dou2017endtoend} \cite{tewari2018high} or both \cite{sanyal2019learning} can be used to regress the parameters of such models in order to best match an image. See also \cite{jackson2017vrn} \cite{zeng2019df2net} \cite{sengupta2018sfsnet}. In addition, some works (e.g.\ \cite{richardson20163d} \cite{tewari2018high} \cite{tran2019towards} \cite{dib2023s2f2}) capture a residual geometry/texture on top of the "best-fit" model parameters. 

\textbf{Multi-view Democratization:}
The more successful multi-view techniques are typically pursued in the context of high-end systems (perhaps, most notably \cite{medusa}), since multiple cameras (often with carefully calibrated extrinsics) are required in order to obtain high-quality results. It is much more difficult to work with uncalibrated camera extrinsics and a single camera (with multiple images or video input). Most of these methods obtain no more than a simple floating face mask (sometimes with eyes), see e.g.\  the photometric stereo approaches in \cite{kemelmacher2011face}\cite{kemelmacher2013internet}\cite{liang2016head}, the structure from motion approaches in \cite{garg2013dense}\cite{shi2014automatic}\cite{ichim2015dynamic}, the optimization approaches in \cite{blanz2003face}\cite{garrido2016reconstruction}\cite{piotraschke2016automated}\cite{roth2016adaptive}, and the neural network approaches in \cite{dou2018multi} \cite{tewari2019fml} \cite{wu2019mvfnet} \cite{bai2020deep} \cite{chaudhuri2020personalized}.
Of these, only \cite{tewari2019fml}\cite{chaudhuri2020personalized} aim to generate de-lit textures (the others either use no texture or simply splat the image onto the geometry). Only \cite{ichim2015dynamic} generates more than a floating face (they generate a full head, eyes, mouth, and animation rig, although the hair and eyebrows are incorrectly flattened into the texture). Most of these works focus on frontal or near-frontal views (occasionally showing obviously distorted side views); in fact, only \cite{liang2016head}\cite{tewari2019fml} show comparisons with ground truth side and/or three-quarters views.

\textbf{Texture:}
In addition to geometry, texture is also acquired in many of the aforementioned approaches; however, there are a number of works that focus primarily on texture acquisition.
\cite{kim2021learning} generate textures with baked-in lighting from a single image by training machine learning models in an unsupervised manner, and \cite{slossberg2022unsupervised} adopts a similar method while removing the baked-in lighting. \cite{smith2020morphable} \cite{han2023learning} aim to build a morphable face albedo/reflectance model by leveraging high quality texture data from high-end capture systems, and \cite{feng2022trust} \cite{ren2023improving}\cite{rainer2023neural} train neural networks to reconstruct textures.

\textbf{Building Animation Rigs:} 
High-end special effects companies have proprietary in-house software, and democratized efforts are primarily limited to various Maya \cite{maya} plugins, Rigify \cite{rigify} in Blender \cite{blender}, Headshot2 from Reallusion \cite{Headshot2}, FaceGen from Daz3D \cite{Daz3D}, and MetaHumans \cite{mhc} in the Unreal Engine \cite{unrealengine}; in particular, Mesh2Metahuman \cite{m2m} is the only publicly available tool that enables the creation of animatable geometry using more than just a single front-facing image. It is also worth noting \cite{shi2020fast} \cite{lin2021meingame}, which train a network to infer animatable models directly. In addition to the identity priors used to regularize neutral/expressionless geometry capture (discussed above), both 3DMM \cite{blanz1999morphable} and FLAME \cite{li2017learning} (and other models) have a separate set of blendshapes meant for facial animation (see \cite{lewis2014practice} for a review). Most of the aforementioned prior works that build animation rigs use these more academic models, perhaps since they are only aiming for technical demonstrations and not (actual) democratization. In real-time applications, joint transforms are often used instead of or in conjunction with blendshapes \cite{ward2004game}. The neutral expression is first deformed with joint-based linear-blend skinning, and blendshapes are (optionally) subsequently used as correctives (see e.g.\ \cite{riglogic22}).

\textbf{Neural Rendering and Implicit Representations:}
Although we aim to build explicit geometry and de-lit textures so that they can be widely used in a variety of existing graphics pipelines, recent and interesting results in avatar generation have used NeRFs \cite{mildenhall2021nerf} and implicit methods (such as \cite{saito2019pifu} \cite{yariv2020idr} \cite{li2022implicit} \cite{alldieck2022photorealistic}) to represent 3D models. See e.g.\ \cite{sevastopolsky2020relightable}\cite{zhang2022fdnerf}\cite{gafni2021nerface} \cite{guo2021adnerf} \cite{wang2021learning} \cite{cao2022authentic} \cite{gao2022reconstructing} \cite{zhang2022physically}\cite{zheng2023pointavatar}\cite{lin2023single}.
Notably, implicit representations can be converted to explicit representations (see e.g.\ \cite{azinovic2023high}\cite{zheng2023neuface}\cite{wang2023sunstage}), although the resulting explicit representations have not yet demonstrated the efficacy required in order to be widely adopted; of particular interest, \cite{wang2023sunstage} is motivated by light stage democratization, but their reconstructed geometry is difficult to evaluate since only frontal facing results are shown.
There have also been attempts at building neural animation rigs, see e.g.\ \cite{10.1145/3588432.3591556}.

%%%%%%%%%%%%%%%%%%%%%%%%%%%%
% Initial Reconstruction
%%%%%%%%%%%%%%%%%%%%%%%%%%%%
\section{Initial Reconstruction}
\label{sec:initialgeo}

We utilize two separate methods for the initial reconstruction of the geometry in order to bootstrap our process. The first method (presented in this section) assumes that one has access to a modern cellphone, and requires (a non-expert user) taking a few pictures from different angles and distances. The second method (see section \ref{sec:video}) is more appropriate when one lacks access to the subject (e.g.\ one might desire a younger version of themself, the subject may be deceased, etc.\! ), and only requires a short video of the subject (e.g.\ from a webcam, YouTube, etc.\! ).

\figcaption{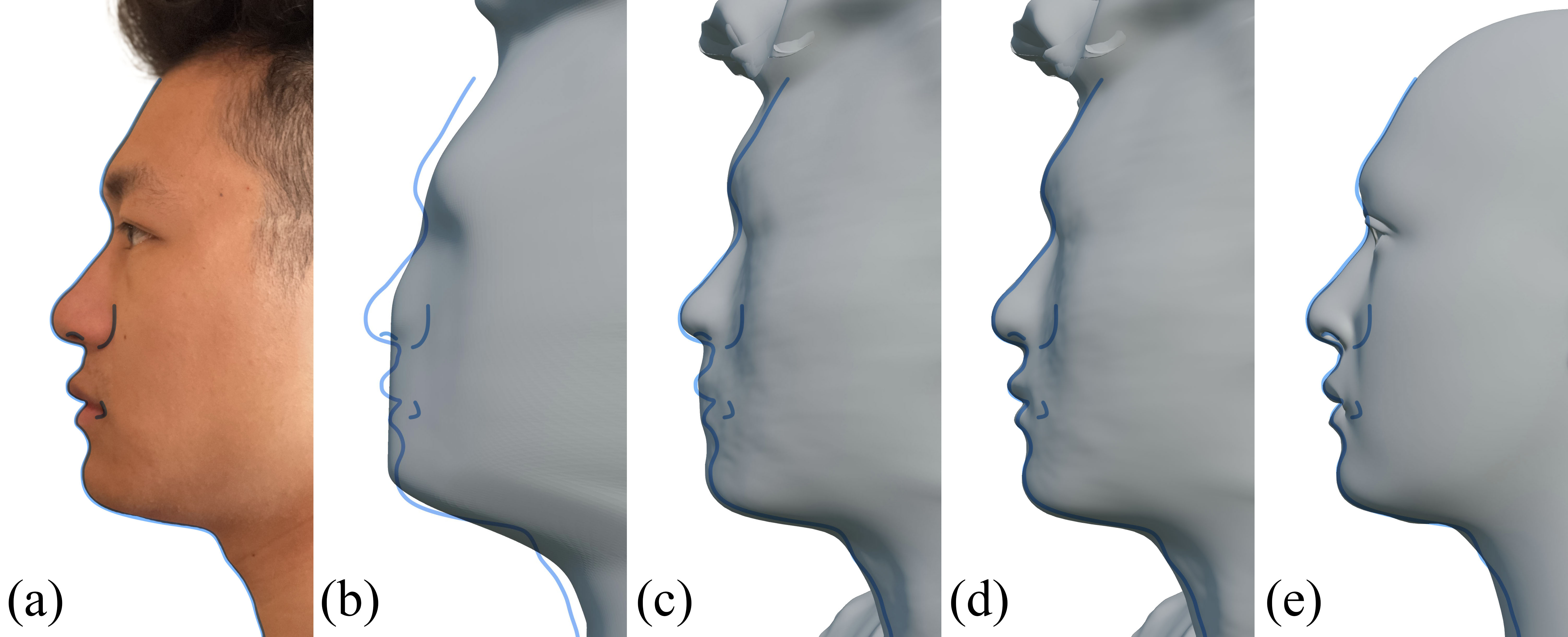}{From left to right: (a) captured image with a (blue) tracing of the silhouette, nostril, and lip corner, (b) initial triangle mesh created from the front view, (c) pixel-aligned projection of the front view triangle mesh onto the rough scan (with the aid of Laplacian smoothing), (d) accounting for silhouette boundaries of adjacent views, (e) MetaHuman reconstruction (note how fitting to a template hallucinates and modifies geometry).}{fig:initial_recon}

We chose an Apple iPhone 13 Pro in order to demonstrate the process (although it is straight forward to extend these techniques to other phones). The back-facing dual camera was used to capture stereo color images from five views (front, left/right three-quarters, and left/right profile). For each view, stereo block matching \cite{konolige1998small} was used to obtain a crude depth estimate; then, a pixel-aligned signed distance function was computed on a voxelized view frustum \cite{rasmussen2003smoke}, the voxelized view frustum was resampled onto a Cartesian grid, the fast marching method \cite{sethian1999fast} was used to obtain signed Euclidean distance on the Cartesian grid, marching tetrahedra \cite{doi1991efficient} was used to construct a triangulated surface mesh, and segmentation \cite{yu2018bisenet} on the color images was used to remove background vertices (not corresponding to the subject). See Figure \ref{fig:initial_recon}(b). To rigidly align the five triangle meshes, landmark detection \cite{hyprsense2020} was run on one image from each view and the point on the triangle mesh corresponding to each visible landmark was identified (via ray tracing); then, the Procrustes algorithm \cite{gower1975generalized} was used to coarsely align view-adjacent pairs of meshes and iterative closest point \cite{holz2015registration} was used to refine this coarse alignment.

In order to refine the five triangle meshes and combine them into a single mesh, we obtained a rough scan from the iPhone front TrueDepth camera. For each view, one of the camera/image pairs was used to perturb the triangle mesh vertices in the pixel-aligned look-direction in order to best match the surface of the rough scan. These perturbations were regularized using Laplacian smoothing \cite{field1988laplacian} on the displacements; importantly, this smoothing helps to alleviate mismatches between the triangle mesh geometry and the rough scan (e.g.\ mismatching nostril vertices with the cheek portion of the scan) while also providing displacements along silhouette boundaries that do not overlap with the scan. See Figure \ref{fig:initial_recon}(c). Afterwards, each mesh was again perturbed in the pixel-aligned look-direction (again with Laplacian smoothing) in order to match the silhouette boundaries of its (one or two) adjacent view(s). See Figure \ref{fig:initial_recon}(d). Finally, screened Poisson surface reconstruction \cite{kazhdan2013screened} was used on the point cloud formed by the combined vertices of all five triangle meshes.

Along the lines of \cite{lin2022leveraging}, we photon map the mesh to obtain a texture suitable for the Mesh2Metahuman pipeline \cite{m2m}. The Mesh2Metahuman pipeline retopoligizes the mesh to be consistent with an underlying MetaHuman animation rig \cite{mhc} (see section \ref{subsec:mhrig}). Since the resulting neutral identity blendshape $N^{\mathrm{mh}}$ may be far from the input mesh (both due to a lack of variety of scanned identities in the dataset and regularization), the Mesh2MetaHuman pipeline also outputs displacements $D^{\mathrm{mh}}$ that perturb the vertices so that $N^{\mathrm{mh}}+D^{\mathrm{mh}}$ is closer to the input mesh. Of course, one could use a different dataset and/or different optimization scheme; in fact, we found that subsequently optimizing to minimize the displacements (with a per-vertex loss) produced a neutral identity blendshape $N$ significantly closer to $N^{\mathrm{mh}}+D^{\mathrm{mh}}$ than $N^{\mathrm{mh}}$ (resulting in an $N+D$ with a smaller $D$).

It is worth noting that a number of prior works can achieve results similar to that shown in Figure \ref{fig:initial_recon}(d) and thus obtain results that are similar to Figure \ref{fig:initial_recon}(e) via the Mesh2Metahuman pipeline (perhaps using \cite{lin2022leveraging} or similar methods to obtain the input texture required to execute the Mesh2Metahuman pipeline). However, regularized and hallucinated geometry will adversely affect the likeness, the Mesh2Metahuman tool does not contain a method for obtaining a de-lit texture, and the default animation rig will not actuate in accordance with the motion signatures of the subject. We address these issues in section \ref{sec:neutralgeo}, \ref{sec:neutraltex}, and \ref{sec:rigmorph} respectively.

%%%%%%%%%%%%%%%%%%%%%%%%%%%%
% Geometry Refinement
%%%%%%%%%%%%%%%%%%%%%%%%%%%%
\section{Geometry Refinement}
\label{sec:neutralgeo}

In a democratized pipeline, camera intrinsics might be available (to some accuracy); however, knowledge of lighting, albedo, and other information required for disentanglement of the geometry from the texture and lighting will be lacking. In particular, a synthetic rendering of the current guess for the geometry will have different features than the real-world image because of mismatches in geometry, texture, and lighting. Our key insight is that one can bake (entangled) lighting and texture information from the real-world image onto the current guess for the geometry in order to provide surrogate landmark information (similar in spirit to painting on mo-cap dots) for subsequent geometry optimization.

The main issue with baking in lighting and texture in order to create surrogate features is that there is a mismatch between the synthetic and real-world geometry. Thus, after an initial rigid alignment, each real-world image is non-linearly warped to better align with the synthetic geometry before projecting the pixel colors from the real-world image into the texture of the synthetic geometry. See Figure \ref{fig:tex_refinement}. Afterwards, the synthetic geometry is optimized to match the original unwarped image with the aid of the surrogate features that have been baked into its texture. Although using multiple views and the appropriate (different) projected texture for each view is essential to the efficacy of this process, we initially describe the method in terms of a single view (in sections \ref{subsec:rigid_alignment_lighting}-\ref{subsec:geometry_opt}) before describing the modifications required to accommodate a multi-view approach (in section \ref{subsec:geometrybootstrap}). 

\figcaptionwide{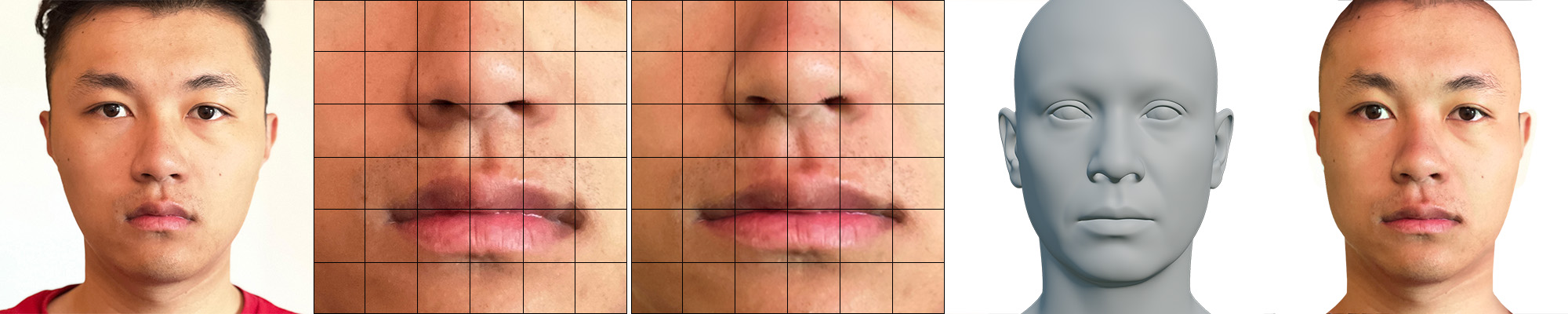}{A real-world image (shown in the first figure) is warped (in image space) to better match a synthetic rendering of a current guess for the geometry (using an appropriate texture). A zoomed-in view of the real-world image is shown before (second figure) and after (third figure) the warp. The fourth figure shows the current geometry, and the fifth figure shows the result obtained by projecting the warped real-world image onto that geometry. This new texture contains baked-in lighting that provides surrogate features useful when optimizing the synthetic geometry to match the original unwarped image.}{fig:tex_refinement}

\subsection{Rigid Alignment and Lighting}
\label{subsec:rigid_alignment_lighting}

Given a current geometry and texture, we use landmarks \cite{hyprsense2020} to rigidly align it with a real-world image; subsequently, a spherical harmonics lighting model \cite{ramamoorthi2001efficient} is optimized to best match a synthetic rendering (of the geometry/texture) to the real-world image. This rigid alignment and estimated lighting is required in order to obtain commensurate results when using a segmentation map network \cite{yu2018bisenet} on both the synthetic rendering and the real-world image (segmentation maps are used in the construction of the non-linear warp).

Given a landmark tracker of choice, we hand-label a (non-sliding) subset of the landmarks on the MetaHuman template geometry (this hand-labeling only needs to be done once per choice of landmark tracker); then, a simple least-squares fit can be used to provide an initial estimate for the rigid alignment between any geometry and any image. Similar to \cite{wu2023deep}, we refine the initial estimate by comparing landmarks computed on renderings of the geometry as opposed to using the hand-labeled landmarks; importantly, this allows landmarks to slide and alleviates issues caused by consistent errors in the landmark tracker, such as always labeling a specific marker too high/low/etc.\ on similar images (these errors are more prevalent for profile angles and out-of-distribution images). Assuming the intrinsic camera parameters are known, we solve for the extrinsic parameters $\mathbf{T}$ (rotation and translation) to refine the rigid alignment by minimizing $$\mathcal{L}_{rigid}(\mathbf{T}) = \sum_{l} \big[\zeta(\Psi(\mathbf{v}, \mathcal{T}, L^a, \mathbf{T}))_l - \zeta(\mathcal{I}^{R})_l \big]^2$$ where $\Psi$ differentiably rasterizes  (we use \cite{ravi2020pytorch3d}) the geometry (vertex positions $\mathbf{v}$, texture $\mathcal{T}$) into an image using (only) ambient lighting $L^a$. $\zeta$ is a landmark tracker that operates on images, $l$ are the computed landmarks (\cite{hyprsense2020} computes 159 landmarks), and $\mathcal{I}^R$ is the real-world image. Note that \cite{hyprsense2020} defines each landmark as a weighted sum of heat-map values in order to preserve differentiability.

Although we found ambient lighting to be sufficient for the landmark tracker, it was insufficient for the segmentation map network from \cite{yu2018bisenet}. Therefore, we next estimate the parameters of a spherical harmonics lighting model $L$ (while keeping $\mathbf{T}$ fixed) by minimizing a per-pixel objective function
$$\mathcal{L}_{light}(L) = \sum_{i \notin \mathcal{B}} \big[\Psi(\mathbf{v}, \mathcal{T}, L, \mathbf{T})_i - \mathcal{I}_i^{R} \big]^2$$ that ignores background pixels.

\subsection{Warping and Projection}
\label{subsec:warping}

A synthetic rendering of the current geometry and texture (using the spherical harmonics lighting estimate) will typically be quite different than the real-world image due to the so-called "domain gap"; moreover, it is difficult (if not impossible) to disentangle the errors in geometry, texture, and lighting from each other. Thus, we use semantic segmentation \cite{yu2018bisenet}  of key coarse regions of the face (e.g.\ nose, lips, eyebrows, silhouette) as the signal for aligning a real-world image with a synthetic rendering. The semantic segmentation also helps to account for subtle differences in expression (alleviating the difficulties associated with capturing expressions with a generic model).

We construct an optical flow field $\mathbf{f}$ to smoothly warp a one-hot encoding of the semantic segmentation of the real-world image $\mathbbm{1}(Seg(\mathcal{I}^R))$ to match (as well as possible) a one-hot encoding of the semantic segmentation of the synthetic rendering $\mathbbm{1}(Seg(\mathcal{I}^S))$.  Note that we merge the hair and body regions into the background category of the one-hot encoding in order to focus on the facial shape. The per-pixel objective function \begin{equation}\mathcal{L}_{seg}(\mathbf{f}) = \sum_{i} \big( \mathbbm{1}(Seg(\mathcal{I}^S))_i - \eta(\mathbbm{1}(Seg(\mathcal{I}^R)), \mathbf{f})_i \big)^2 \end{equation} compares $\mathbbm{1}(Seg(\mathcal{I}^S))$ to a non-linear warp of $\mathbbm{1}(Seg(\mathcal{I}^R))$  where $\eta$ warps one image to another using the flow field $\mathbf{f}$ and bilinear interpolation. Since $\mathcal{L}_{seg}$ only provides penalties near differences in the segmentations, we add regularization to minimize the per-pixel norm of $\mathbf{f}$ and the per-pixel Laplacian (using a 5x5 stencil) of $\mathbf{f}$.

Although using $\mathcal{L}_{seg}$ and the regularizers behaved as expected, we found that post-processing $\mathbf{f}$ led to significantly improved results. The post-process solves a 2D Laplace equation (using a standard 5-point stencil) using Dirichlet boundary conditions on pixels with differently-labeled neighbors (according to $Seg(\mathcal{I}^R)$) and Neumann boundary conditions on the image boundary. This preserves $\mathbf{f}$ on the boundaries of $Seg(\mathcal{I}^R)$ while guaranteeing that it is smooth elsewhere.

Lastly, we project the warped (and thus better aligned) real-world image $\mathcal{I}^W$ into the texture for the synthetic geometry using the photon mapping technique proposed in \cite{lin2022leveraging}. Each pixel of $\mathcal{I}^W$ is projected onto the geometry to determine texture coordinates for storing a "photon" based on the pixel color. In the gathering step, the color for each pixel in the texture map is determined using a weighted average of the $k$ nearest photons. Typically, the weights would decay with distance; however, we additionally scale the weights with a power law for specular falloff $(\mathrm{max}(0,-\mathbf{n} \cdot \mathbf{r}))^p$ where $\mathbf{n}$ is the (unit) normal to the geometry and $\mathbf{r}$ is the (unit) ray direction from the pixel to the geometry. This additional scaling diminishes the contribution from photons near occlusion boundaries, which may still be misaligned after the warp.

\subsection{Optimization}
\label{subsec:geometry_opt}

Since the texture $\mathcal{T}$ computed via section \ref{subsec:warping} already contains baked-in lighting, only ambient lighting $L^a$ is required for the differentiable rasterizer $\Psi$ when using inverse rendering to optimize the geometry. Perturbations of the (geometry) vertex positions $\mathbf{v}$ can be computed by minimizing a per-pixel objective function $$\mathcal{L}_{photo}(\mathbf{v}) =  \sum_{i \notin \mathcal{B}} \big[ \Psi(\mathbf{v}, \mathcal{T}, L^a, \mathbf{T})_i - \mathcal{I}^R_{i}  \big]^2$$ that ignores background pixels (note that the hair and body regions are still kept merged into the background). 

In order to encourage the matching of semantic information about facial shape, we additionally utilize a second objective function based on the semantic segmentations of $\mathcal{I}^R$ and the synthetically rendered $\mathcal{I}^S$. Let $\mathbf{v}^{k}$ be the geometry at the beginning of the $k$-th iteration and $\mathcal{I}^{S,k}$ be the synthetic rendering of that geometry; then, the procedure from section \ref{subsec:warping} can be used to construct an optical flow $\mathbf{f}^k$ that smoothly warps $\mathbbm{1}(Seg(\mathcal{I}^R))$ to $\mathbbm{1}(Seg(\mathcal{I}^{S,k}))$.
Given sample points $\mathbf{v}^k_j$ on the geometry $\mathbf{v}^k$, a ray tracer can be used to compute their screen-space locations $\Tilde{\Psi}(\mathbf{v}^k_j, \mathbf{T})$; then, the per-geometry-sample objective function $$\mathcal{L}^k_{sem}(\mathbf{v}) =  \sum_{j} \big[ \Tilde{\Psi}(\mathbf{v}_j, \mathbf{T}) - \Tilde{\Psi}(\mathbf{v}^k_j, \mathbf{T}) + \mathbf{f}^k_{j})  \big]^2$$ compares the screen-space motion of each sample point to the optical flow $\mathbf{f}^k$ that smoothly warps $\mathbbm{1}(Seg(\mathcal{I}^{S,k}))$ to $\mathbbm{1}(Seg(\mathcal{I}^R))$ at that location ($\mathbf{f}^k_{j}$ is the interpolation of $\mathbf{f}^k$ to the screen-space location $\Tilde{\Psi}(\mathbf{v}^k_j, \mathbf{T})$). Although there are many potential choices for the sample points (e.g.\ one could choose all visible triangle vertices), we use the sub-triangle locations that rasterize to the center of each non-background pixel (removing the need to interpolate $\mathbf{f}^k$). This makes the sample points $\mathbf{v}_j$ depend on the triangle vertices via barycentric interpolation.

Since the lighting is baked into the surrogate texture, neither $\mathcal{L}_{photo}$ nor $\mathcal{L}^k_{sem}$ has or requires any dependence on the normal vectors of $\mathbf{v}$; thus, geometric regularization is essential. We utilize two geometric regularizers: $\mathcal{L}_{edge}$ and $\mathcal{L}_{laplace}$.  $\mathcal{L}_{edge}$ penalizes changes in the edge lengths, computed by comparing edge lengths in the current mesh to the lengths of the corresponding edges in the initial pre-optimization mesh. $\mathcal{L}_{laplace}$ penalizes the Laplacian (using a one-ring stencil) of the vertex displacements, computed by comparing vertex positions in the current mesh to their corresponding positions in the pre-optimization mesh; importantly, this promotes smooth displacements, but does not smooth out features in the initial pre-optimization mesh as the Laplacian of the vertex positions would.

\subsection{Using Multiple Views}
\label{subsec:geometrybootstrap}

Since both rigid alignment and segmentation maps struggle with non-front-facing views (especially with non-photorealistic textures), we begin the multi-view process by warping and projecting the front-facing real-world image into the texture (as per subsection \ref{subsec:warping}); then, this photorealistic texture is used to compute rigid alignment and segmentation maps for each of the other views. Although one might blend textures from the various views together into a single texture, this incorrectly averages lighting information (which varies according to view); thus, we compute (as per subsection \ref{subsec:warping}) and maintain separate textures for each view, and subsequently optimize one geometry with separate losses ($\mathcal{L}_{photo}$ and $\mathcal{L}^k_{sem}$, each using the view-appropriate $\mathcal{T}$) for each view combined into the same objective function (noting that one could optionally increase the weights on $\mathcal{L}_{edge}$  and $\mathcal{L}_{laplace}$ in order to balance the increased forces from using multiple views). In addition, the rigid alignment and segmentation maps seemed to perform best when using the texture from the previous iteration corresponding to the view under consideration (when that texture exists).

The entire multi-view approach can be repeated iteratively, obtaining a new geometry after each iteration. We obtained the best results using only a single front-facing view in the first iteration and multiple views in subsequent iterations. Since this iterative approach jointly optimizes both the geometry and the camera extrinsics (for the utilized real-world images), it can be beneficial to repeat the initial image-based reconstruction of the geometry (in section \ref{sec:initialgeo}) using these improved camera extrinsics; notably, we found that this led to a significant improvement in both the initial image-based reconstruction as well as the subsequent multi-view iteration (which one might expect given the efficacy of end-to-end approaches). See Figure \ref{fig:refined_recon}.

\figcaption{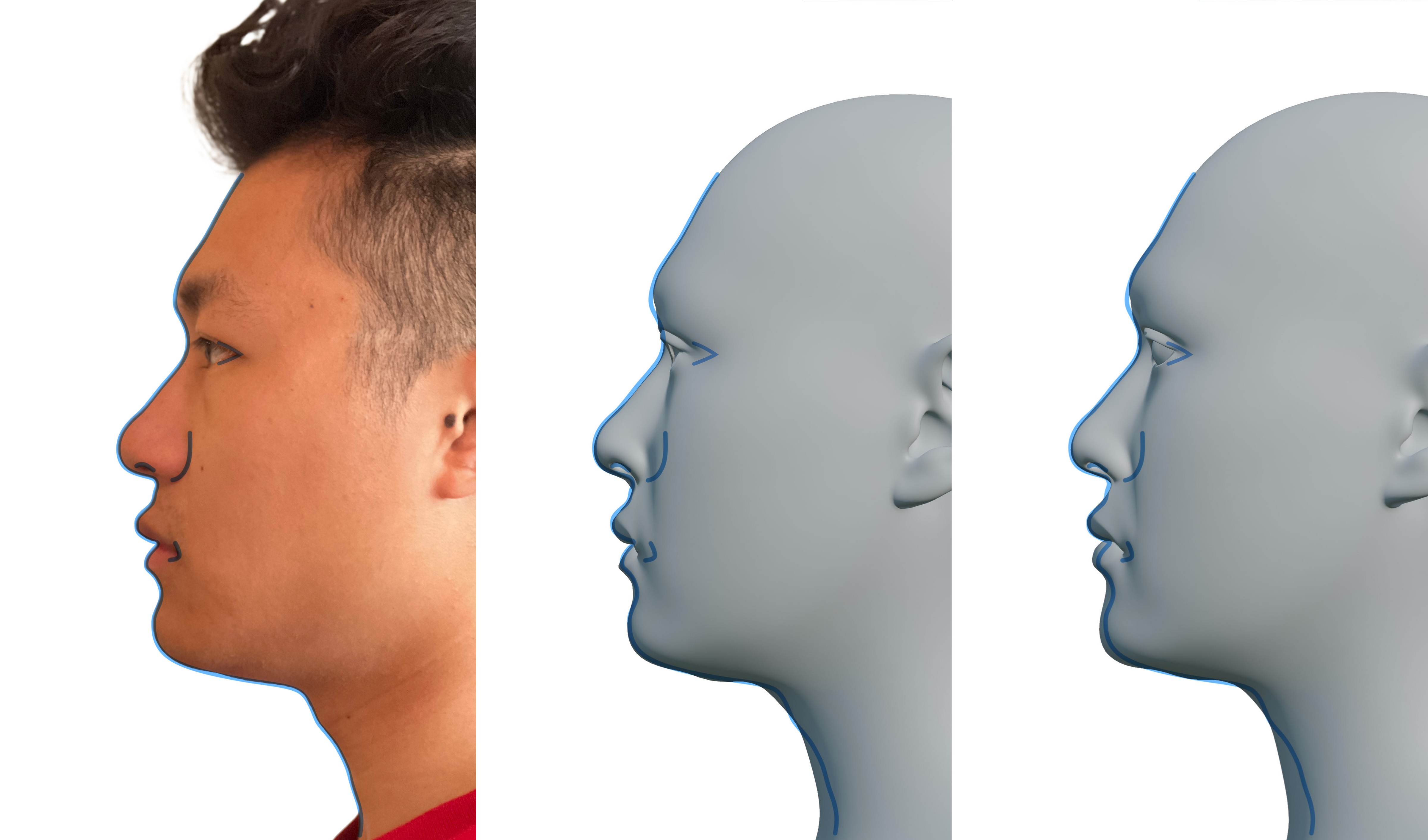}{From left to right: (a) captured image with a (blue) tracing of the silhouette, nostril, lip corner, eye corner, and mouth corner, (b) MetaHuman reconstruction from section \ref{sec:initialgeo} Figure \ref{fig:initial_recon}, (c) the results obtained by using our geometry refinement process (in section \ref{sec:neutralgeo}). In particular, note the improvements in the eye and mouth regions.}{fig:refined_recon}

%%%%%%%%% NEUTRAL TEX
\section{De-Lighting Texture}
\label{sec:neutraltex}

The real-world entanglement of geometry, texture, and lighting coupled with human perception makes it difficult to ascertain the quality of a reconstruction (this is especially misleading in prior works that compare an entangled reconstruction to a reference image); however, disentangled results are required in order to use a reconstruction in a standard graphics pipeline (especially when the geometry/texture needs to be re-lit). See Figure \ref{fig:floating_face}.

\figcaption{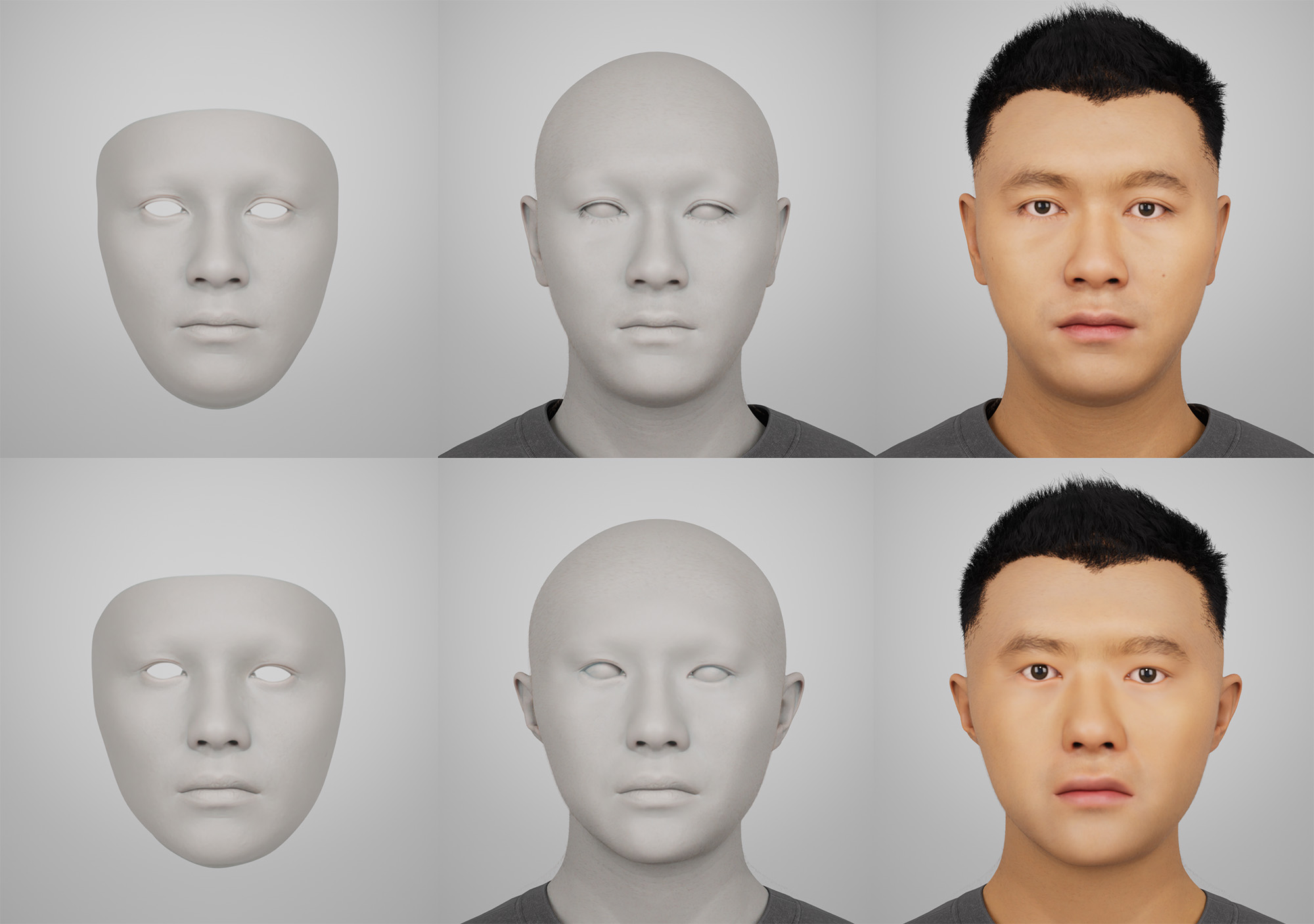}{It is often hard to distinguish the efficacy of a floating, grey-shaded face mask (first column). Adding the head, neck, eyes, etc.\ gives a better indication of an identity (second column). However, only after adding textures and hair does human perception cause the images to start migrating into and through the uncanny valley (third column). The top row shows a state-of-the-art geometry reconstruction result from the Mesh2MetaHuman toolset, and the bottom row shows our reconstructed geometry. The Mesh2MetaHuman reconstruction requires an iPhone 12's depth sensor, which is significantly better than the depth sensors in the iPhone 13 and newer models (the Mesh2MetaHuman reconstruction quality is remarkably worse on the iPhone 13 and newer models). The third column shows our textures, hair, and eyebrows on both reconstructed geometries, since the Mesh2MetaHuman toolset lacks the ability to reconstruct textures disentangled from lighting and geometry.}{fig:floating_face}

An initial guess for the texture can obtained from the warping and projection discussed in section \ref{sec:neutralgeo}, the photon mapping algorithm discussed in \cite{lin2022leveraging}, or any other prior work; however, most methods result in misaligned textures due to imperfections in the predicted geometry.
Thus in section \ref{subsec:texavg}, we discuss how to leverage the methods proposed in section \ref{sec:neutralgeo} in order to obtain better texture alignment.
In section \ref{subsec:levlight}, we illustrate how lighting estimates can be used to remove baked in lighting via inverse rendering.
In section \ref{subsec:freqsep}, we discuss the separation of an acquired texture into high-frequency and low-frequency components and the subsequent projection of the low-frequency component into a pre-computed PCA basis (further removing baked in lighting information).

\subsection{Texture Alignment and Averaging}
\label{subsec:texavg}

Importantly, popular face representation frameworks (we use the Metahuman \cite{mhc}) contain at least one default texture (often synthetic) that has been correctly aligned with the texture coordinates on the geometry; thus, we aim to align our subject-specific texture with that default texture.
At the end of the geometry refinement (in section \ref{sec:neutralgeo}), each real-world image can be projected into the texture map (and gathered to pixels/texels) in order to obtain a corresponding texture. These projected textures will typically have stretching artifacts and misalignment in areas where the camera look direction is at a grazing angle to the geometry. Aiming to eliminate these artifacts, we combine all the projected textures into a single stitched texture by choosing the color at each texel (texture map pixel) from the texture map that had the most orthogonal geometry (as compared to the camera look direction) at that texel. Afterwards, the stitched texture is warped to align with the default Metahuman texture; then, all of the projected textures are warped to match the aligned stitched texture. See Figure \ref{fig:tex_align}.

\figcaption{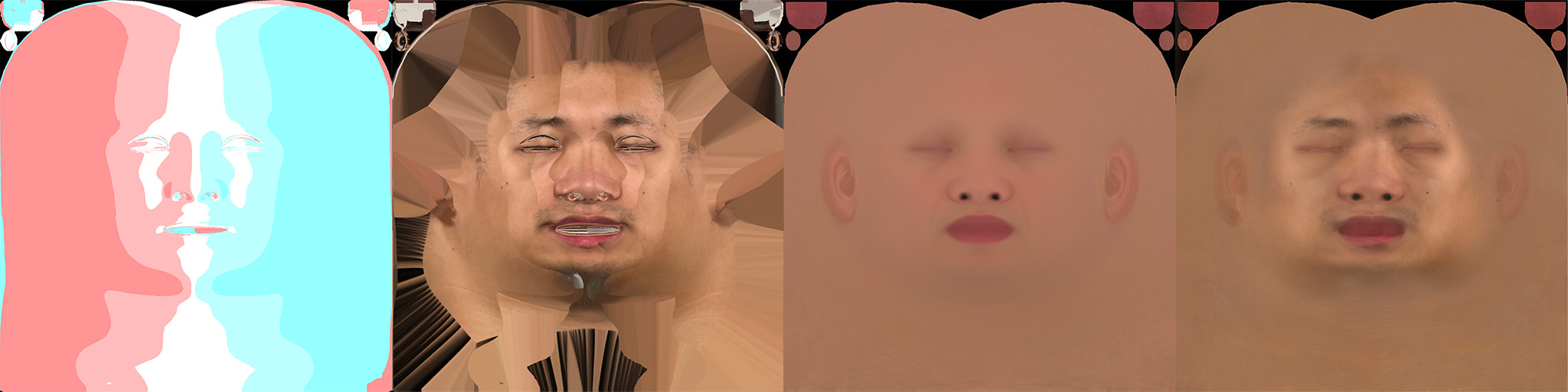}{The first figure shows color-coding to indicate which texture (front view: white, three-quarters view: light red/blue, side view: dark red/blue) is contributing to the stitched texture (in the second figure). The stitched texture is warped to align with the default MetaHuman texture (shown in the third figure), and then each of the individual textures is warped to align with the stitched texture; afterwards, the textures are averaged together using the blending discussed in section \ref{subsec:texavg}. The final result is shown in the fourth figure.}{fig:tex_align}

The warping (in texture space) is accomplished by computing a flow field $\mathbf{f}$ along the lines of section \ref{subsec:warping} with a few modifications. The segmentation map network can be run on a rendered image (using the texture under consideration) and subsequently projected back onto the geometry and into texture space; alternatively, the segmentation map network can be run directly on the texture image (and works surprisingly well).
Since each real-world image only fills in a portion of the full texture, we found it beneficial to include an additional term
$$\mathcal{L}_{lmk}(\mathbf{f}) = \sum_{l} \big( \zeta^{\mathrm{mh}}_l - \eta(\zeta(\mathcal{I}^R), \mathbf{f})_l \big)^2$$
that encourages landmark alignment. $\zeta^{\mathrm{mh}}_l$ are the fixed locations in texture space of hand-labeled landmarks on the MetaHuman template.
The landmark locations $\zeta(\mathcal{I}^R)$ can be computed on the real-world image $\mathcal{I}^R$ and subsequently projected into texture space or be computed directly on the projected texture (similar to the segmentation map).
$\eta$ warps the landmarks using bilinear interpolation (to the pre-warped sub-pixel landmark locations) of the flow field $\mathbf{f}$.

When warping each of the individual textures to align with the stitched texture, we utilize (dark) moles as additional landmarks (in the above equation). A square sample window is used to identify texels that are darker than the mean texel color in the window. The center of each connected component is identified as a mole whenever the connected component's area is larger than a threshold.
Mole correspondence (between the stitched texture and the texture that is being warped to match it) is established by a greedy algorithm that finds the closest mole with similar area working outwards from the tip of the nose.

After warping, the textures are combined together using a specular falloff weighting along the lines of that discussed at the end of section \ref{subsec:warping}.  
Notably, since the face was rotated in a static lighting environment, this averaging helps to remove baked in lighting information.
We additionally reduce the per-texel specular falloff weighting of a texture that has an adjacent view that is more orthogonal to the geometry at that texel, since the adjacent view can likely provide a better approximation to the texture.
For each view, orthogonality is measured at each texel as $-\mathbf{n} \cdot \mathbf{r}$; importantly, this orthogonality measure is labeled to be unusable whenever it is negative. Using this measure of orthogonality, the specular falloff weighting is further multiplied by the ratio of the orthogonality measures between the current texture and an adjacent texture whenever that ratio is less than one (when there are two choices for the adjacent texture, the one with the smaller ratio is used).

\subsection{Leveraging Lighting Estimates}
\label{subsec:levlight}

An approximation to the real-world lighting can be used to improve upon the result from section \ref{subsec:texavg}. This can be achieved numerically using inverse rendering (e.g.\ to estimate a spherical harmonics approximation \cite{ramamoorthi2001efficient}) or captured directly (e.g.\ via a chrome sphere environment map). Some phone apps (e.g. HDReye \cite{hdreye}) use HDR capture and automatic stitching to create environment maps; in addition, consumer-level $360^\circ$ cameras (e.g.\ Insta360 X3 \cite{insta360x3} and Ricoh Theta \cite{ricohtheta}) are becoming more prevalent. We have also experimented with programming an at-home large-screen TV in order to mimic various light patterns from a light stage (see also \cite{sengupta2021light}).

Given any of the aforementioned lighting estimates, we can improve the texture from section \ref{subsec:texavg} via inverse rendering. The objective function includes the per-pixel differences between the real-world image and the corresponding synthetic rendering for each of the five views, but these per-pixel differences are given specular falloff weightings 
(discussed at the end of section \ref{subsec:warping})
in order to diminish contributions along the silhouette boundaries of each view.
A Laplacian term is used on the texel deltas to ensure smooth changes in the texture. The spherical harmonics parameters can be jointly optimized, if desired.

\subsection{Frequency Separation and PCA Projection}
\label{subsec:freqsep}

Whether a texture is acquired via the algorithms proposed in section \ref{subsec:texavg} or 
\ref{subsec:levlight} or the photon mapping process from section \ref{sec:neutralgeo} (or other previous work), the result needs to be converted into a fully de-lit albedo texture (required in a standard graphics pipeline). 
At this point in our process, the majority of the baked lighting information is contained in the lower frequencies of the texture, while personalized features such as moles tend to be contained in the higher frequencies. 
Thus, we separate the texture into high-frequency and a low-frequency components by applying a Gaussian low-pass filter. 

We remove the baked-in lighting from the low-frequency component via PCA projection. Starting with the MetaHuman database of albedo textures (from real-world scans), we perform whitening and subsequently calculate a PCA basis. Then, we optimize the coefficients of the first five bases in order to match the pixel color of the whitened low-frequency component using an L2 regularizer.
We ignore eyebrow regions (since they are incorrectly baked into the texture), eye folds/sockets, and the inner nostril during PCA projection, but do reconstruct these regions.
In order to capture some of the facial stubble, moles, and other details, the high-frequency component is added back to the non-ignored regions. See Figure \ref{fig:texture_lf_hf}.

\figcaption{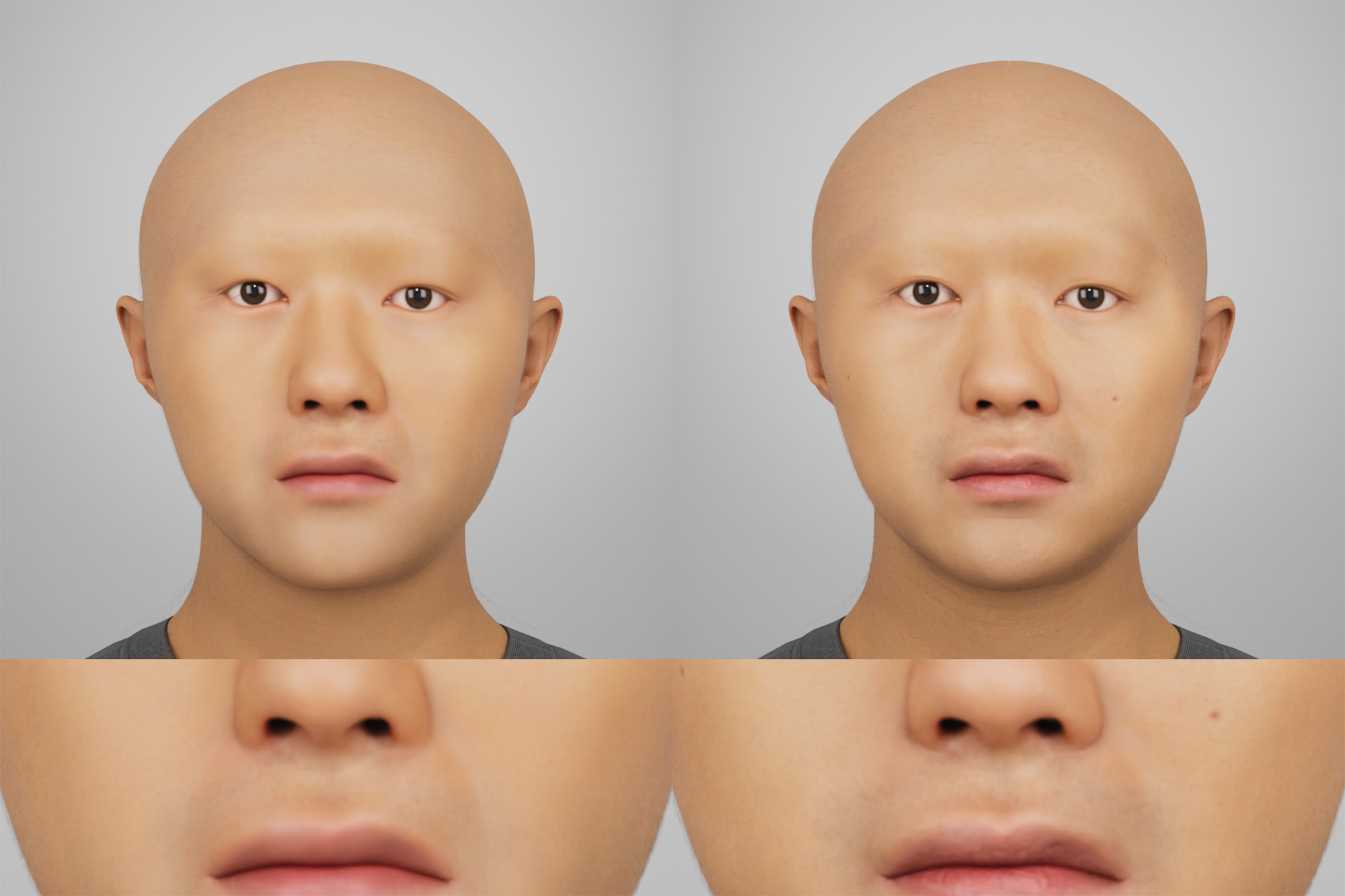}{PCA-projected low frequency texture (left). Final texture (right), after adding back the high-frequency component. Bottom row is a zoomed-in version of the top row. Note the facial stubble, lip wrinkles, and mole.}{fig:texture_lf_hf}

We select the best-fit hair and eyebrows from the MetaHuman preset options. Since the MetaHuman database has relatively limited choices for hair and eyebrows (which one would expect to improve over time given the variety of character customization abilities in recent high-end video games), we slightly (manually) sculpt the hair and eyebrows to better match the images. The character customization controls for the eyes are more readily usable (no manual sculpting is necessary). Notably, character customization options are becoming more democratized, and non-expert users (non-artists) can reasonably match hair, eyebrows, etc.\ (when appropriate options are available); however, a non-expert user (or even a trained artist) would struggle to capture the overall face shape and appearance (which is the focus of Sections \ref{sec:neutralgeo} and \ref{sec:neutraltex}). See Figures \ref{fig:yilin_comparison}, \ref{fig:dalton_comparison}, and \ref{fig:lookdev}.

\figcaption{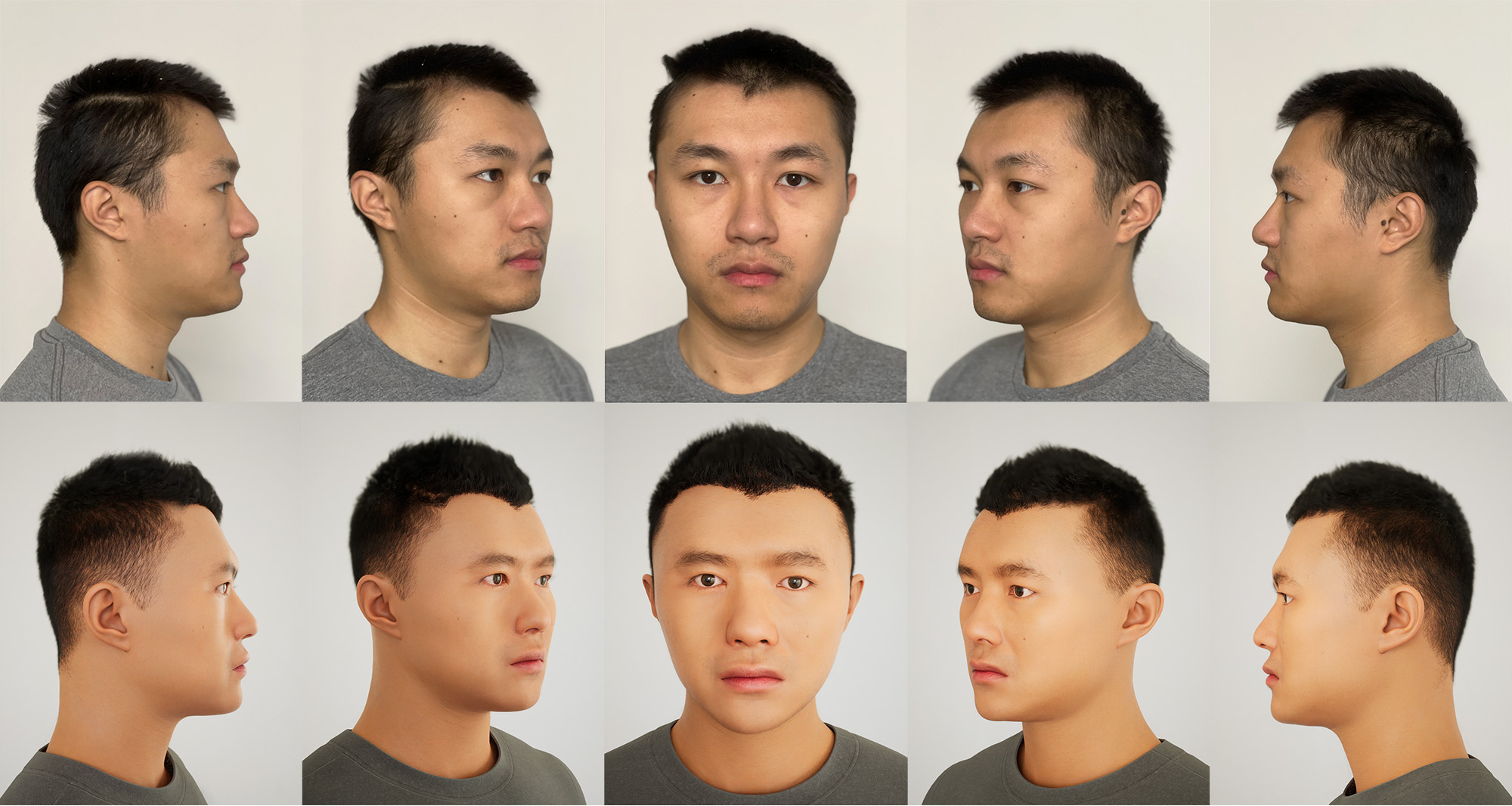}{Comparison between captured images and our reconstructed geometry/texture for five views. The synthetic rendering utilizes lighting similar to that present in the real-world images. See also Figure \ref{fig:lookdev}.}{fig:yilin_comparison}

\figcaption{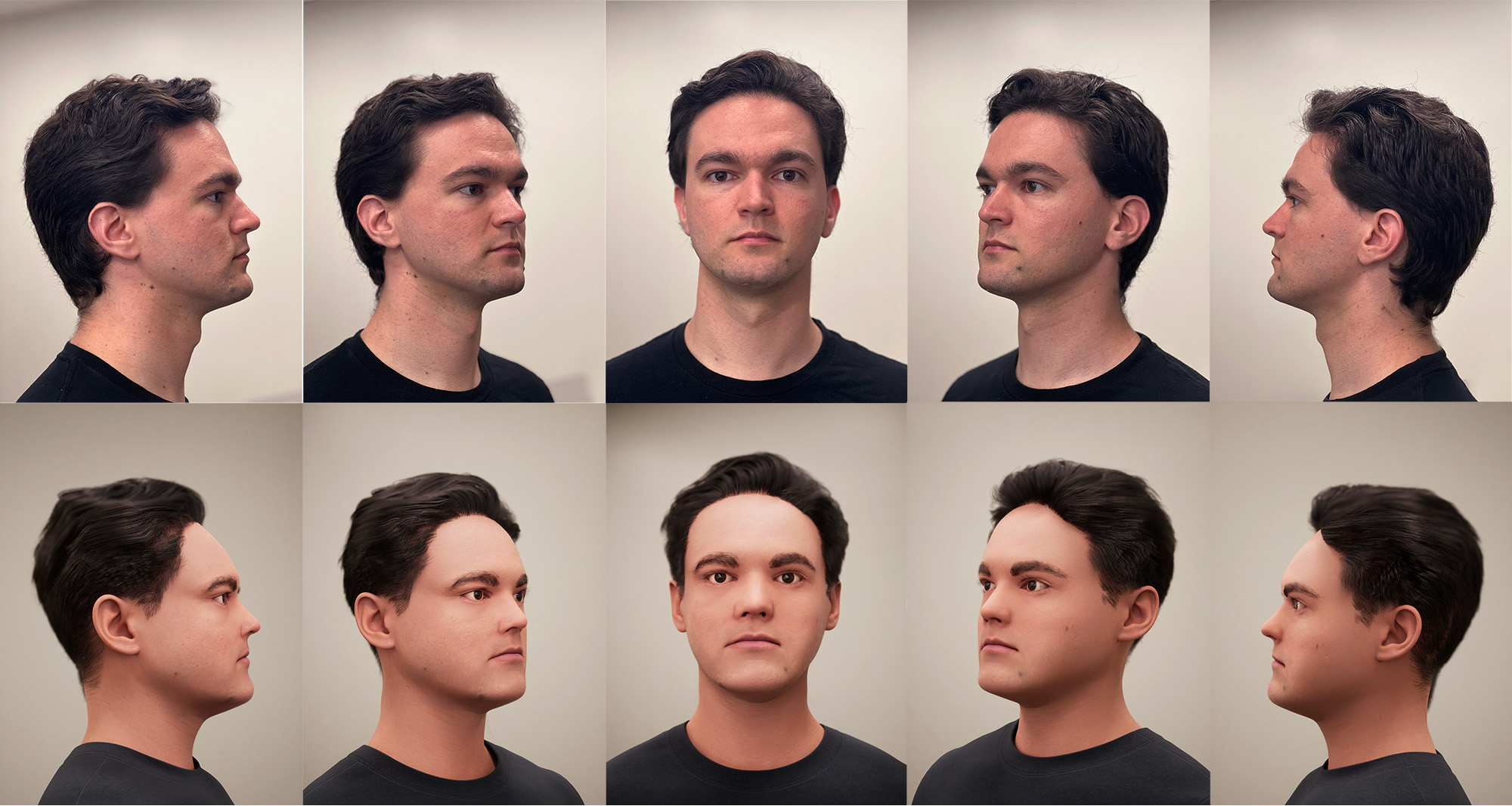}{Comparison between captured images and our reconstructed geometry/texture for five views. The synthetic rendering utilizes lighting similar to that present in the real-world images. See also Figure \ref{fig:lookdev}.}{fig:dalton_comparison}

\figcaption{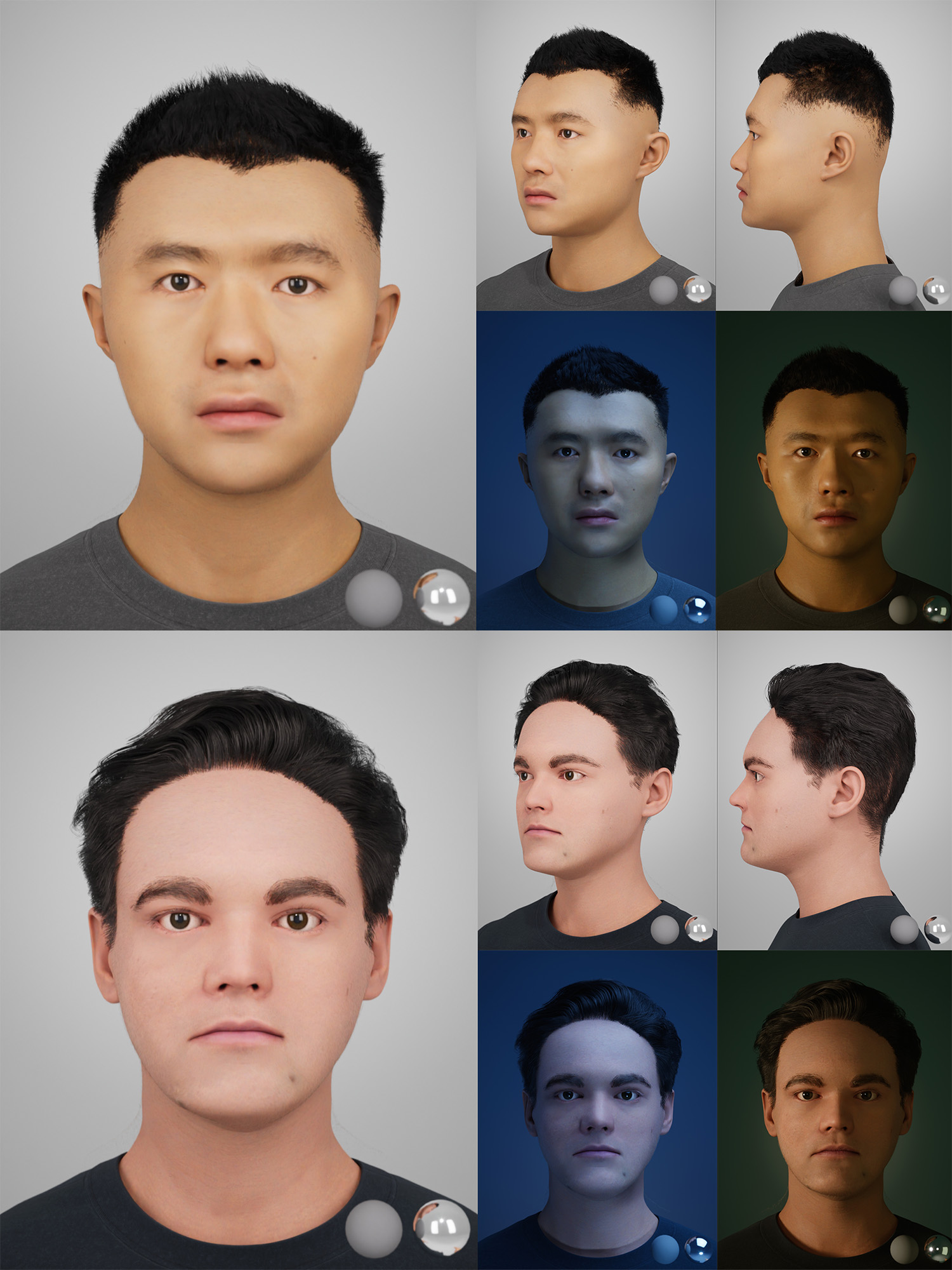}{The reconstructed avatars from Figures \ref{fig:yilin_comparison} and \ref{fig:dalton_comparison} rendered under various novel lighting conditions.}{fig:lookdev}

\section{Video-based Reconstruction}
\label{sec:video}

In this section, we briefly discuss how our method adapts to the situation when one lacks access to the subject (e.g.\ one might desire a younger version of themself, the subject may be deceased, etc.\! ). When high-quality images of a neutral expression in each of our five required views are available (either as images or as a subset of a webcam/YouTube/etc.\ video), the methods proposed in sections \ref{sec:neutralgeo} and \ref{sec:neutraltex} require no modification; however, the initial reconstruction (from section \ref{sec:initialgeo}) would need to be replaced with template MetaHuman geometry (typically degrading the quality of the final result). A better initial guess could be obtained by training a deepfake model on the video footage (as discussed in \cite{lin2022leveraging}). When images of a neutral expression in each of our five required views are not available, a deepfake model (trained on the video footage) could be used to create surrogate images suitable for our pipeline. Figure \ref{fig:mark} illustrates the results we obtained on two publicly recognizable figures.

\figcaption{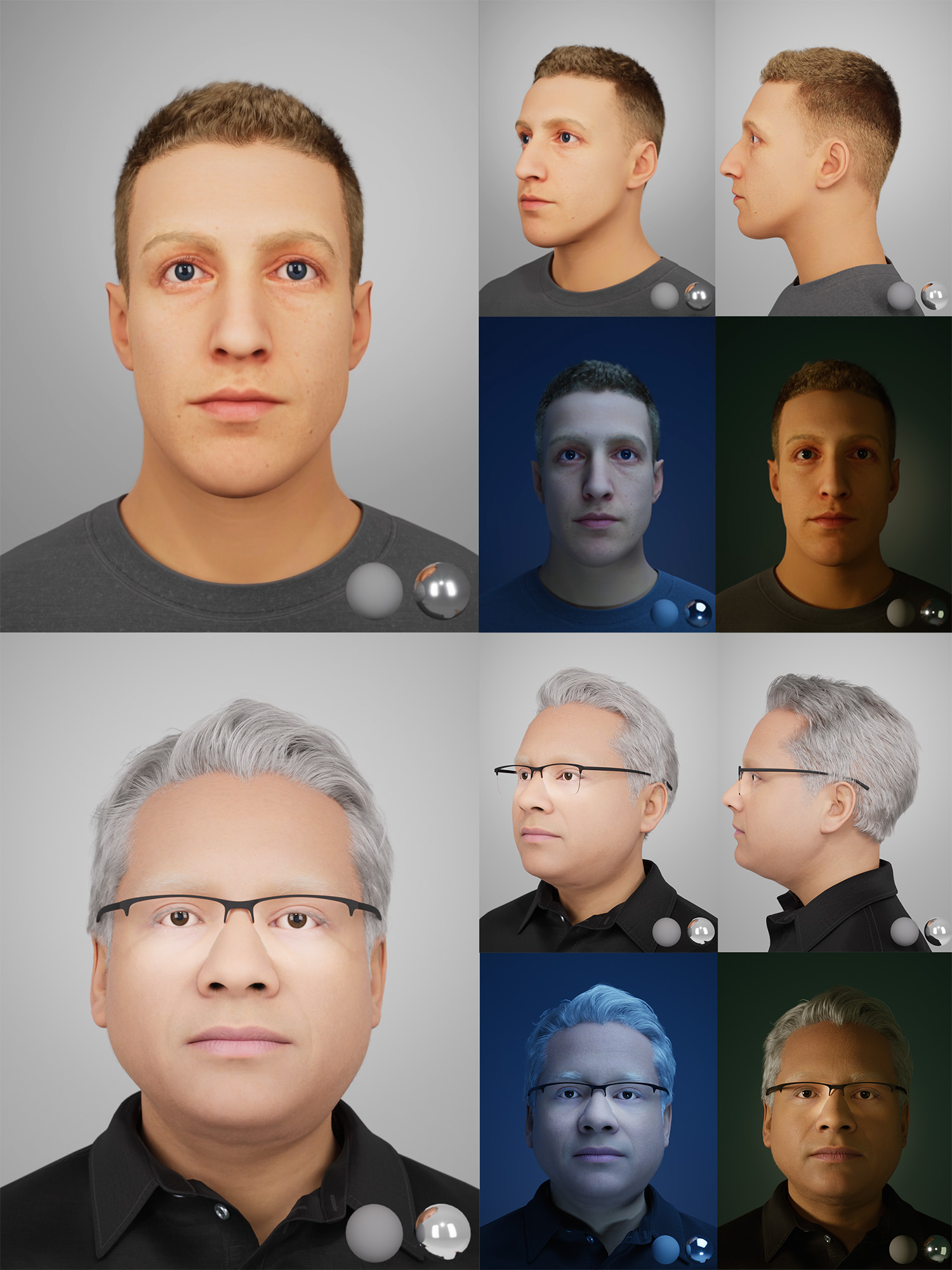}{In order to demonstrate the efficacy of our approach (from sections \ref{sec:neutralgeo} and \ref{sec:neutraltex}) when one does not have access to the subject, we chose two publicly recognizable figures to reconstruct from online images/video.}{fig:mark}
%%%%%%%%% RIG MORPH
\section{Animation Rig}
\label{sec:rigmorph}

The Mesh2Metahuman pipeline creates geometry topoligized to be consistent with an underlying animation rig (see section \ref{subsec:mhrig}). 
As noted near the end of Section \ref{sec:initialgeo}, additional displacements (on top of the neutral identity blendshape) are typically required in order to better match an input mesh. 
When these additional displacements are large (see e.g.\ Figure \ref{fig:volmorph}), subsequent animations often possess undesirable artifacts. Section \ref{subsec:morph} discusses how a volumetric morph of the animation rig removes the need for these additional displacements. Although the morphed animation rig fits the reconstructed geometry quite well, it still will not have the same motion signatures as the real-world subject (e.g.\ the subject's actual smile will be dissimilar to a dialed-in smile on the animation rig). We address this by capturing a number of basic expressions via a Simon Says approach (Section \ref{subsec:simonsays}), and subsequently using them to build a personalized animation rig (Section \ref{subsec:rigbuild}).

\subsection{MetaHuman Animation Rig}
\label{subsec:mhrig}

The MetaHuman animation rig \cite{riglogic22} is freely available in the Unreal Engine \cite{unrealengine}, and the open-source code can be used outside the Unreal Engine as well. The top level implementation exposes a set of sliders that non-experts can readily use to control a face, see Figure \ref{fig:2dGUI}. For the sake of real-time applications, deformations from a neutral geometry are primarily implemented via joints and linear blend skinning. The approximately 200 control sliders are mapped to translations and rotations of about 800 joints. A more computationally expensive version of the animation rig uses about 700 blendshape correctives on top of the joint-based deformations.

\figcaption{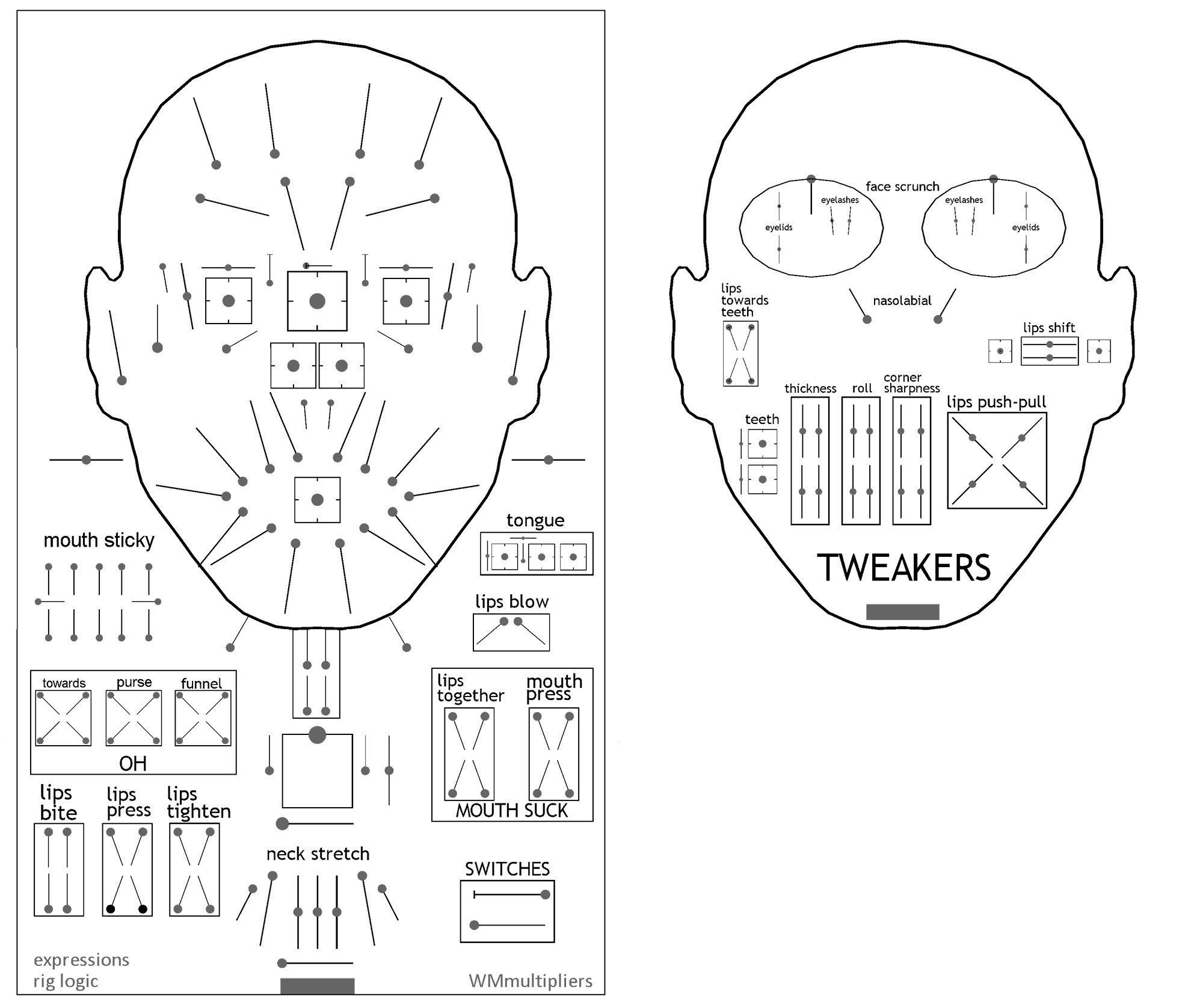}{2D graphical user interface for MetaHuman animation rig control.}{fig:2dGUI}

Given a textured face geometry (the texture needs to be suitable for landmark detection), the Mesh2MetaHuman pipeline \cite{m2m} creates an appropriately topologized neutral identity blendshape $N^{\mathrm{mh}}$ as well as a corresponding animation rig. The placements of the joints, the mappings from slider controls to joint displacements (position and orientation), the linear blend skinning weights, the blendshape correctives, and the mappings from slider controls to blendshape corrective weights are all determined automatically by leveraging a database of expertly hand-crafted animation rigs. As is typical, this database was constructed from light stage (and other) scans of various individuals. Due to limited representability in the database, $N^{\mathrm{mh}}$ can deviate quite a bit from the input geometry; thus, an additional displacement blendshape $D^{\mathrm{mh}}$ is added on top of $N^{\mathrm{mh}}$. See Figure \ref{fig:volmorph}. Although $D^{\mathrm{mh}}$ mostly rectifies the inability of the database to adequately represent the input geometry, the animation rig is meant for $N^{\mathrm{mh}}$ not $N^{\mathrm{mh}}+D^{\mathrm{mh}}$. See Figure \ref{fig:volmorph1}. This can lead to various animation artifacts (we remedy this in section \ref{subsec:morph}). See Figure \ref{fig:volmorph2}. It is also important to note that one would not generally expect the blending of animation rigs from a database to adequately capture the motion signatures of any particular individual not already in that database (we address this in Sections \ref{subsec:simonsays} and \ref{subsec:rigbuild}).

\subsection{Volumetric Morph}
\label{subsec:morph}

Figure \ref{fig:volmorph} shows input geometry that is not well-represented by the MetaHuman database. This leads to a reconstructed neutral blendshape $N^{\mathrm{mh}}$ that differs significantly from the input geometry; thus, additional displacements $D^{\mathrm{mh}}$ are added on top of $N^{\mathrm{mh}}$. Problematically, the animation rig is intended for $N^{\mathrm{mh}}$ not $N^{\mathrm{mh}}+D^{\mathrm{mh}}$. See Figures \ref{fig:volmorph1} and \ref{fig:volmorph2}. We remedy this by volumetrically morphing the animation rig to better fit $N^{\mathrm{mh}}+D^{\mathrm{mh}}$, allowing for the removal of the extra displacements entirely. In particular, we morph from $N^{\mathrm{mh}}$ to $\hat{N}$, where $\hat{N}$ represents the result of our geometry optimization (from Section \ref{sec:neutralgeo}).

\figcaption{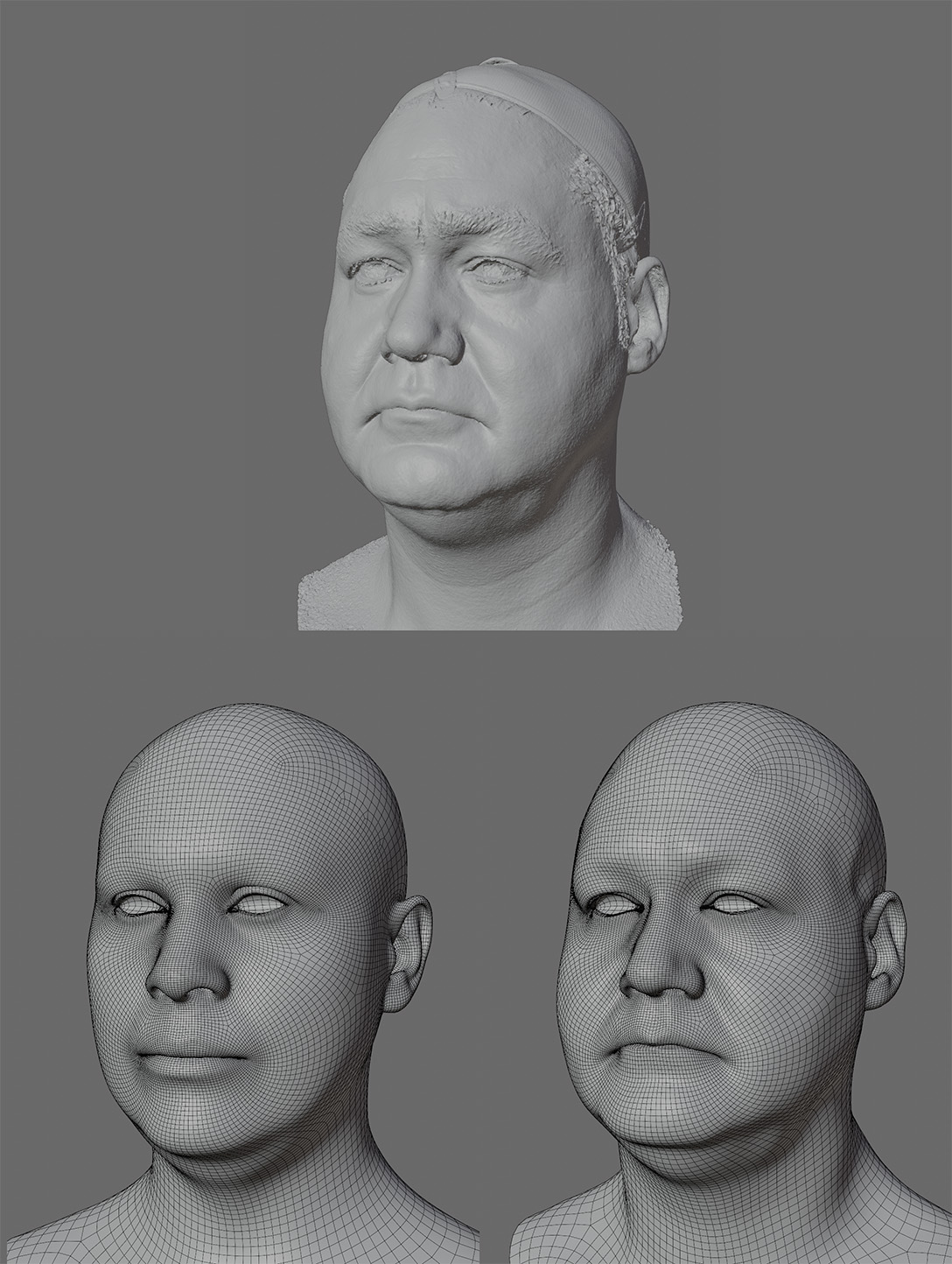}{Top: scan of an individual that is not well-represented by the MetaHuman database. Bottom left: identity blendshape $N^{\mathrm{mh}}$ reconstructed from the MetaHuman database. Bottom right: adding an additional displacement blendshape $D^{\mathrm{mh}}$ to the identity blendshape $N^{\mathrm{mh}}$ results in a better match to the scan.}{fig:volmorph}

\figcaption{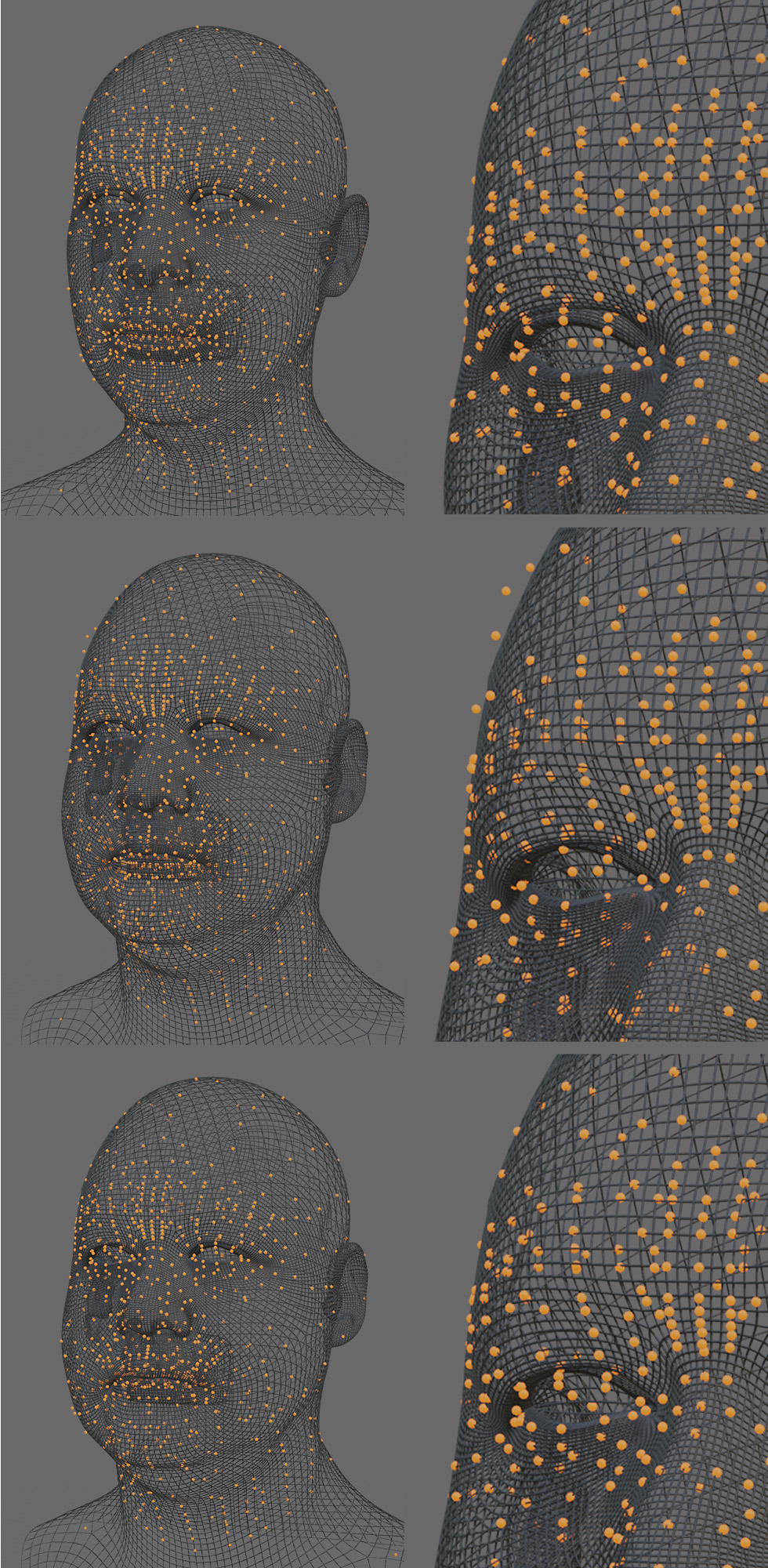}{Top: the Mesh2MetaHuman animation rig fits $N^{\mathrm{mh}}$ well, by design. Middle: the Mesh2MetaHuman animation rig does not fit $N^{\mathrm{mh}}+D^{\mathrm{mh}}$ well. Joints can be too deep inside of or even outside of the surface geometry; perhaps more importantly, joints are improperly aligned with the surface topology. Bottom: our volumetrically morphed animation rig fits $N^{\mathrm{mh}}+D^{\mathrm{mh}}$ well.}{fig:volmorph1}

\figcaption{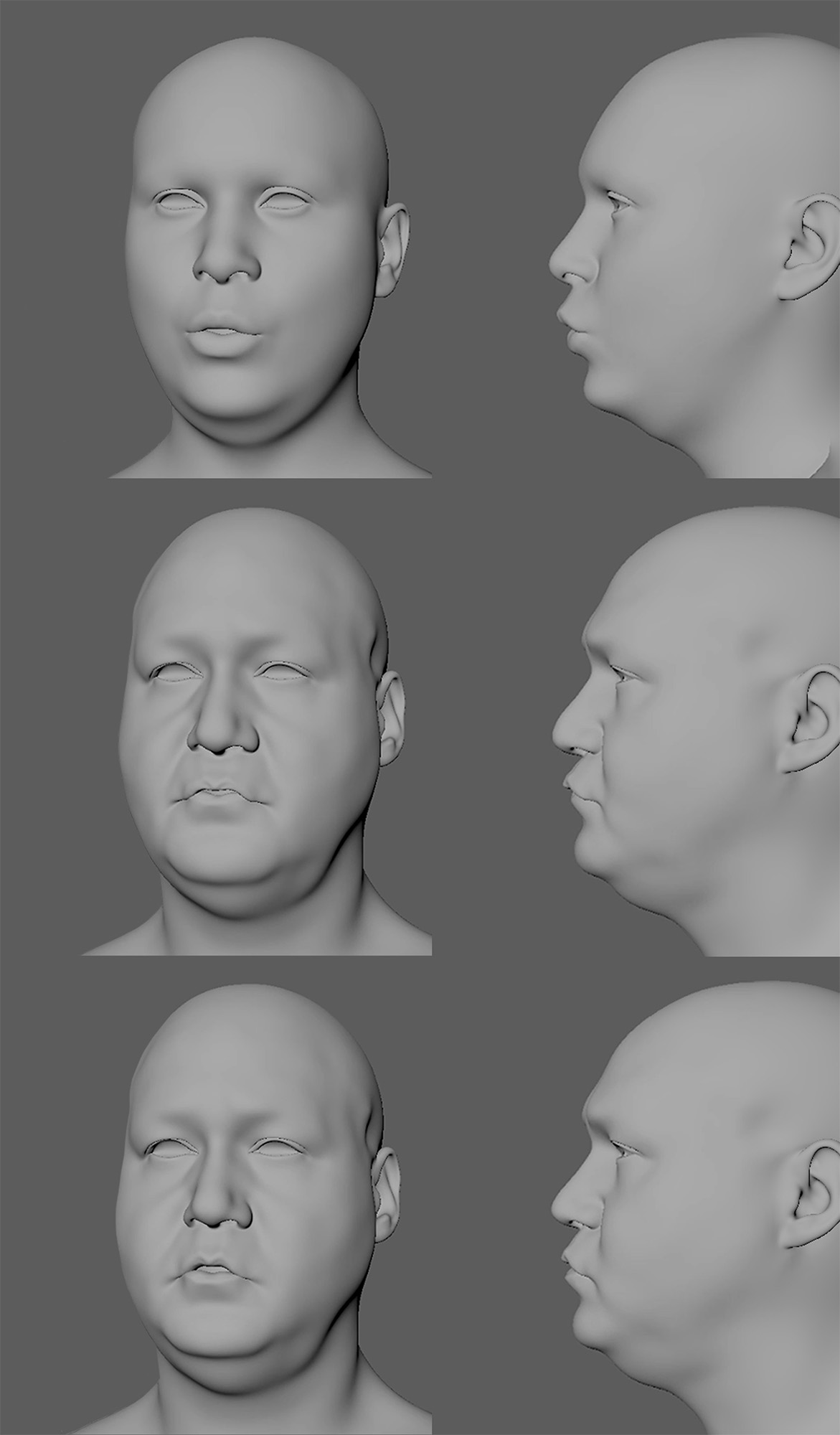}{Top row: the animation rig fits $N^{\mathrm{mh}}$ well, and also animates it well. Middle row: The animation rig does not fit $N^{\mathrm{mh}}+D^{\mathrm{mh}}$ well, and thus does not animate it well. Bottom row: our morphed animation rig fits $N^{\mathrm{mh}}+D^{\mathrm{mh}}$ well, and also animates it well. All three rows have identical animation rig controls (for purse and funnel) dialed in. Note the semantically similar expressions in the top and bottom rows (as expected); in contrast, the poor-fitting animation rig (middle row) incorrectly makes a semantically dissimilar expression.}{fig:volmorph2}

Following \cite{cong2015fully}, we extend the $N^{\mathrm{mh}}$ to $\hat{N}$ per-vertex displacements to a volumetric field by solving three decoupled three-dimensional Poisson equations. The computational domain is specified by an oversized bounding box and discretized with a Cartesian grid of suitable resolution. Dirichlet boundary conditions are specified on any Cartesian grid edge that intersects $N^{\mathrm{mh}}$ using barycentric interpolation of the per-vertex displacements. After solving the Poisson equations, trilinear interpolation can be used to determine the displacements required to morph each joint's position. In order to update a joint's orientation, three additional sample points ($x_0+\epsilon \mathrm{i}$, $x_0+\epsilon \mathrm{j}$, and $x_0+\epsilon \mathrm{k}$ where $x_0$ is the unmorphed joint position and $\epsilon$ is a small number) are morphed to their new locations and subsequently used to determine a rotation (taking care to re-orthogonalize after the morph). The blendshape correctives can be rewritten in the local coordinates of $N^{\mathrm{mh}}$, identity-transformed to the local coordinates of $\hat{N}$, and then transformed back out of local coordinates (independent of the volumetric morph). The mappings from slider controls to joint displacements (position and orientation), the linear blend skinning weights, and the mappings from slider controls to blendshape corrective weights are left unchanged.

\subsection{Simon Says} 
\label{subsec:simonsays}

One would not generally expect the blending of animation rigs from a database to adequately capture the motion signatures of any particular individual not already in that database; thus, we capture a number of basic expressions and subsequently use them to build a personalized animation rig (see Section \ref{subsec:rigbuild}). To balance between animation rig quality and capture session length (and complexity), we select 27 expressions (divided into two groups):
\begin{itemize}
    \item pucker, nose wrinkle, cheek raise, mouth stretch, squint lower eyelid, lip corner pull, jaw open, brow lower, brow raise, blink
    \item  upper lip raise, nose wrinkle combined with upper lip raise, sharp lip corner pull, mouth dimple, lip corner depress, lower lip depress, purse, lips towards, funnel, funnel purse, funnel towards, oh, jaw open extreme, lip corner pull combined with jaw open, mouth stretch combined with jaw open, smile, smile stretch combined with jaw open
\end{itemize}
noting that the second group can (optionally) be omitted for brevity. 

The capture session is guided via a Simon Says approach, where the user's avatar (built via Sections \ref{sec:initialgeo}, \ref{sec:neutralgeo}, \ref{sec:neutraltex}, and \ref{subsec:morph})
makes an expression that the user attempts to match. See Figure \ref{fig:simonsays}.
Instead of giving users obscure terminology for various expressions, visual cues better enable non-experts to quickly understand each expression (especially when the visual cues are on an avatar similar in appearance to the user). The visual interface shows a video feed of the user side by side with their avatar, see Figure \ref{fig:simonsaysimage}.

\figcaption{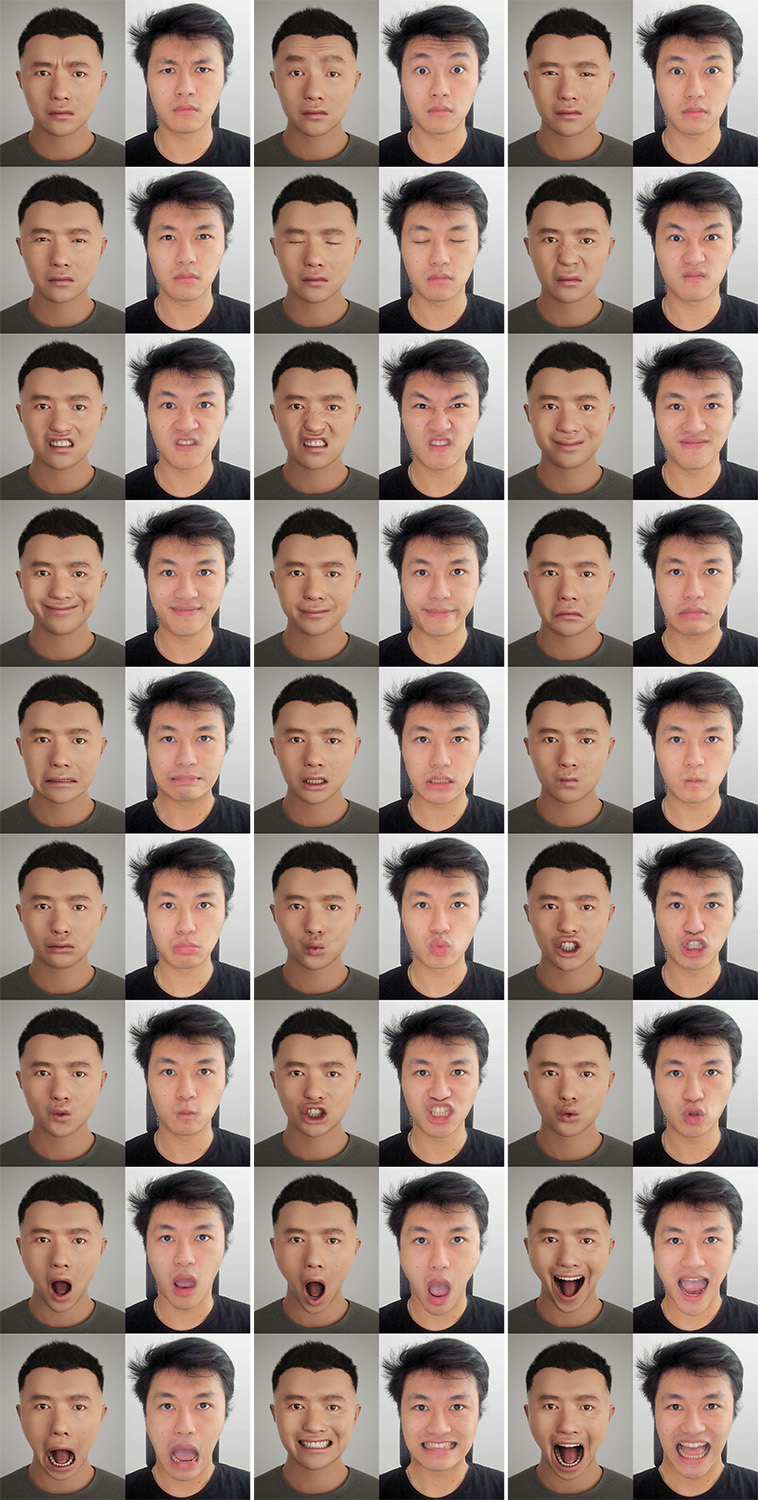}{The 27 expressions we capture in order to build a personalized animation rig. Left: the avatar makes an expression, which is shown to the user (see Figure \ref{fig:simonsaysimage}). Right: the user makes their version of that expression. The differences between the animation rig expressions and the user's expressions are quite apparent; in particular, the user opens their jaw differently, makes different mouth shapes, and sometimes struggles to make a similar expression (for example, see the last column of the first row).}{fig:simonsays}

\figcaption{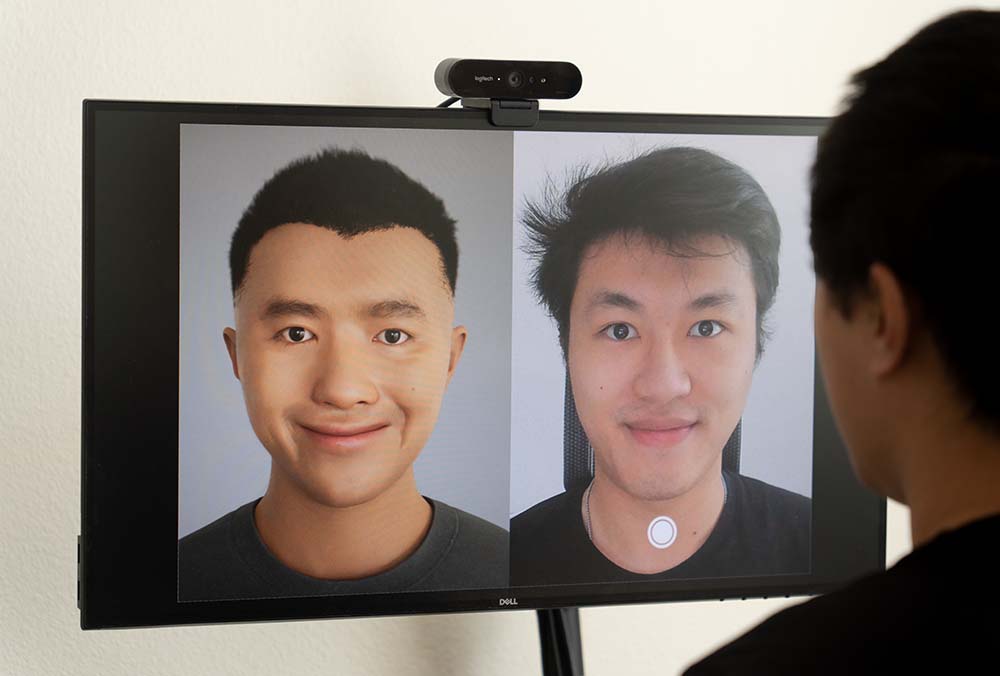}{Visual interface for the Simon Says capture session. A video feed of the user is shown side by side with their avatar. When a pose is finalized, the user clicks a button in order to capture/save the image of their face.}{fig:simonsaysimage}

\subsection{Personalized Animation Rigs}
\label{subsec:rigbuild}

For each of the 27 images from the Simon Says capture session (see Figure \ref{fig:simonsays}, columns 2,4, and 6), the geometry refinement method from Section \ref{sec:neutralgeo} can be used to reconstruct geometry corresponding to the expression. Since only a single image (not multi-view) is available, more regularization is required; thus, we optimize for animation rig degrees of freedom instead of per-vertex displacements (but otherwise follow Section \ref{sec:neutralgeo}). Using animation rig degrees of freedom also allows one to ignore or remove spurious geometry matching unrelated to a given expression (for example, only mouth degrees of freedom should be activated during a smile). Degrees of freedom that should not be active for a given expression can either be ignored during the optimization or be set identically to zero as a post process (after the optimization). In addition, this allows the capture of symmetric expressions (such as a smile) to be used to obtain geometry for asymmetric expressions (such as a left-only or right-only smile) simply by setting (approximately) half of the controls to zero as a post process.

For each of the 27 basic expressions, the reconstructed geometry is used to modify the corresponding expression in the animation rig. Since each of these is a primary expression in the animation rig, they are straightforward to modify. After dialing the sliders corresponding to an expression to their maximum value, the difference between the deformed avatar and the reconstructed geometry is used to replace the corrective blendshape associated with that expression. Optionally, one could instead modify the joint displacements in order to be more consistent with the reconstructed geometry; although, a (smaller magnitude) corrective blendshape would still be required in order to exactly match the reconstructed geometry. Modifying primary expressions also indirectly modifies the complex expressions that are composed of multiple primary expressions. The complex expressions can be preserved by modifying their corrective blendshapes with displacements opposite of those that were used to modify any primary expression that they depend on. A similar approach could be taken when choosing to modify joint displacements. See Figure \ref{fig:rigbuild}for two samples (chosen from the 27, for brevity).

\figcaption{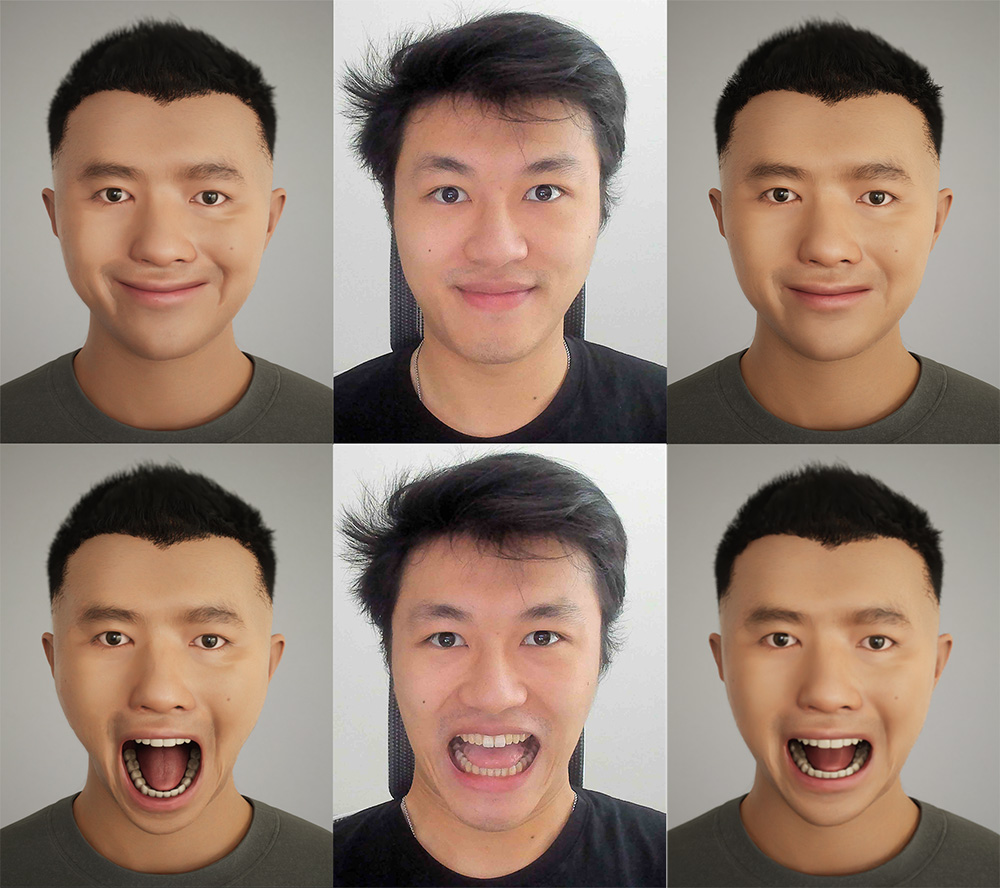}{Before (left) and after (right) modifying the animation rig with reconstructed geometry from the Simon Says capture session (the corresponding image from the Simon Says capture session is shown in the middle).}{fig:rigbuild}

\section{Discussion and Conclusions}
\label{sec:conclusion}

We chose to use the MetaHuman framework because it provides a full head, face, and neck (not just a floating face mask) as well as a default animation rig (however, we did have to devise our own method for adding textures). The use of any other suitable database (either existing or future) will be subject to issues similar to those discussed in this paper: the size of the database will be limited by the difficulties associated with scanning individuals and creating suitable personalized animation rigs, the representability of the database will be lacking due to its limited size (especially when considering the large variance in human faces and expressions), the limited representability will lead to poor results (poor geometry, texture, and animation rigs) for individuals that are not well-represented by the database, etc. Our methods for geometry refinement (Section \ref{sec:neutralgeo}), creating de-lit textures (Section \ref{sec:neutraltex}), morphing template animation rigs to better fit optimized geometry (Section \ref{subsec:morph}), and modifying animation rigs to contain personalized motion signatures (Sections \ref{subsec:simonsays} and \ref{subsec:rigbuild}) should thus be useful for improving the results from any database (not only the MetaHuman database)

We would like to stress the importance of capturing and utilizing profile views, as they are crucial for obtaining the correct geometry (not discernible from front facing views, due to depth ambiguities). Although there is a plethora of work that focuses on primarily front facing views, the importance of capturing and evaluating the results from novel views is becoming more prevalent in the literature, e.g.\ \cite{wu2023lpff}. The importance of obtaining the correct geometry (and texture) becomes even more apparent when subsequently animating/deforming the face; although the neutral identity may appear correct, expressions can lie in the uncanny valley. This is perhaps even more important when one considers subsequent biomechanical simulations where areas and volumes (not only two-dimensional projections) need to be accurate.

Our approach of using baked-in lighting to create surrogate features can be used to improve an initial reconstruction from any method. More importantly, it can be used to improve avatars (such as MetaHumans) directly. This is essential since any mesh-to-rig pipeline will be subject to the inadequacies of its database (and thus both hallucinate and be lossy). Our method for obtaining de-lit textures allows the reconstructed avatar to appear realistic in novel lighting environments while still maintaining the high-frequency moles, stubble, etc.\ required to preserve likeness. Our video-based reconstruction results illustrate that we can create avatars for subjects we do not have access to with only a small modification to our pipeline. 
Finally, our rig building approach alleviates some of the issues associated with the representability gap between any particular individual and a database of (preferably, carefully handcrafted) rigged avatars. The volumetric morph better fits the animation rig to the geometry (removing both deformation artifacts and issues with semantics), and the Simon Says approach enables a non-expert user to capture expressions important for reproducing their motion signatures.

\section{Acknowledgements}
\label{sec:ack}

Research supported in part by ONR N00014-19-1-2285, ONR N00014-21-1-2771, and Epic Games. We would like to thank Reza and Behzad at ONR for supporting our efforts into machine learning.
We would also like to thank Dragan Davidovic, Iain Matthews, Jihun Yu, Jovan Mijatov, Kamy Leach, Kim Libreri, Michael Lentine, Relja Ljubobratović, Steven Caulkin, Thibaut Weise, Jungwoon Park, and Vladimir Mastilovic for their insightful discussions.

\bibliographystyle{ACM-Reference-Format}
\bibliography{main}


\begin{thebibliography}{112}

%%% the Web address.

\ifx \showCODEN    \undefined \def \showCODEN     #1{\unskip}     \fi
\ifx \showDOI      \undefined \def \showDOI       #1{#1}\fi
\ifx \showISBNx    \undefined \def \showISBNx     #1{\unskip}     \fi
\ifx \showISBNxiii \undefined \def \showISBNxiii  #1{\unskip}     \fi
\ifx \showISSN     \undefined \def \showISSN      #1{\unskip}     \fi
\ifx \showLCCN     \undefined \def \showLCCN      #1{\unskip}     \fi
\ifx \shownote     \undefined \def \shownote      #1{#1}          \fi
\ifx \showarticletitle \undefined \def \showarticletitle #1{#1}   \fi
\ifx \showURL      \undefined \def \showURL       {\relax}        \fi
% invisible to TeX
\providecommand\bibfield[2]{#2}
\providecommand\bibinfo[2]{#2}
\providecommand\natexlab[1]{#1}
\providecommand\showeprint[2][]{arXiv:#2}

\bibitem[Alldieck et~al\mbox{.}(2022)]%
        {alldieck2022photorealistic}
\bibfield{author}{\bibinfo{person}{Thiemo Alldieck}, \bibinfo{person}{Mihai Zanfir}, {and} \bibinfo{person}{Cristian Sminchisescu}.} \bibinfo{year}{2022}\natexlab{}.
\newblock \showarticletitle{Photorealistic monocular 3d reconstruction of humans wearing clothing}. In \bibinfo{booktitle}{\emph{Proceedings of the IEEE/CVF Conference on Computer Vision and Pattern Recognition}}. \bibinfo{pages}{1506--1515}.
\newblock


\bibitem[{Autodesk, INC.}(2024)]%
        {maya}
\bibfield{author}{\bibinfo{person}{{Autodesk, INC.}}} \bibinfo{year}{2024}\natexlab{}.
\newblock \bibinfo{booktitle}{\emph{Maya}}.
\newblock
\urldef\tempurl%
\url{https:/ autodesk.com/maya}
\showURL{%
\tempurl}


\bibitem[Azinovi{\'c} et~al\mbox{.}(2023)]%
        {azinovic2023high}
\bibfield{author}{\bibinfo{person}{Dejan Azinovi{\'c}}, \bibinfo{person}{Olivier Maury}, \bibinfo{person}{Christophe Hery}, \bibinfo{person}{Matthias Nie{\ss}ner}, {and} \bibinfo{person}{Justus Thies}.} \bibinfo{year}{2023}\natexlab{}.
\newblock \showarticletitle{High-res facial appearance capture from polarized smartphone images}. In \bibinfo{booktitle}{\emph{Proceedings of the IEEE/CVF Conference on Computer Vision and Pattern Recognition}}. \bibinfo{pages}{16836--16846}.
\newblock


\bibitem[Bai et~al\mbox{.}(2020)]%
        {bai2020deep}
\bibfield{author}{\bibinfo{person}{Ziqian Bai}, \bibinfo{person}{Zhaopeng Cui}, \bibinfo{person}{Jamal~Ahmed Rahim}, \bibinfo{person}{Xiaoming Liu}, {and} \bibinfo{person}{Ping Tan}.} \bibinfo{year}{2020}\natexlab{}.
\newblock \showarticletitle{Deep Facial Non-Rigid Multi-View Stereo}. In \bibinfo{booktitle}{\emph{Proceedings of the IEEE/CVF Conference on Computer Vision and Pattern Recognition}}. \bibinfo{pages}{5850--5860}.
\newblock


\bibitem[Beeler et~al\mbox{.}(2010)]%
        {beeler2010high}
\bibfield{author}{\bibinfo{person}{Thabo Beeler}, \bibinfo{person}{Bernd Bickel}, \bibinfo{person}{Paul Beardsley}, \bibinfo{person}{Bob Sumner}, {and} \bibinfo{person}{Markus Gross}.} \bibinfo{year}{2010}\natexlab{}.
\newblock \showarticletitle{High-Quality Single-Shot Capture of Facial Geometry}. In \bibinfo{booktitle}{\emph{ACM SIGGRAPH 2010 Papers}} (Los Angeles, California) \emph{(\bibinfo{series}{SIGGRAPH '10})}. \bibinfo{publisher}{Association for Computing Machinery}, \bibinfo{address}{New York, NY, USA}, Article \bibinfo{articleno}{40}, \bibinfo{numpages}{9}~pages.
\newblock
\showISBNx{9781450302104}
\urldef\tempurl%
\url{https://doi.org/10.1145/1833349.1778777}
\showDOI{\tempurl}


\bibitem[Beeler et~al\mbox{.}(2022)]%
        {medusa}
\bibfield{author}{\bibinfo{person}{Thabo Beeler}, \bibinfo{person}{Markus Gross}, \bibinfo{person}{Paulo Gotardo}, \bibinfo{person}{Jérémy Riviere}, {and} \bibinfo{person}{Derek Bradley}.} \bibinfo{year}{2022}\natexlab{}.
\newblock \bibinfo{title}{Medusa Facial Capture System}.
\newblock
\newblock
\urldef\tempurl%
\url{https://studios.disneyresearch.com/medusa}
\showURL{%
\tempurl}


\bibitem[Blanz and Vetter(1999)]%
        {blanz1999morphable}
\bibfield{author}{\bibinfo{person}{Volker Blanz} {and} \bibinfo{person}{Thomas Vetter}.} \bibinfo{year}{1999}\natexlab{}.
\newblock \showarticletitle{A morphable model for the synthesis of 3D faces}. In \bibinfo{booktitle}{\emph{Proceedings of the 26th annual conference on Computer graphics and interactive techniques}}. \bibinfo{pages}{187--194}.
\newblock


\bibitem[Blanz and Vetter(2003)]%
        {blanz2003face}
\bibfield{author}{\bibinfo{person}{Volker Blanz} {and} \bibinfo{person}{Thomas Vetter}.} \bibinfo{year}{2003}\natexlab{}.
\newblock \showarticletitle{Face recognition based on fitting a 3D morphable model}.
\newblock \bibinfo{journal}{\emph{IEEE Transactions on pattern analysis and machine intelligence}} \bibinfo{volume}{25}, \bibinfo{number}{9} (\bibinfo{year}{2003}), \bibinfo{pages}{1063--1074}.
\newblock


\bibitem[{Blender}(2023)]%
        {rigify}
\bibfield{author}{\bibinfo{person}{{Blender}}.} \bibinfo{year}{2023}\natexlab{}.
\newblock \bibinfo{booktitle}{\emph{Rigify}}.
\newblock
\urldef\tempurl%
\url{https://docs.blender.org/manual/en/latest/addons/rigging/rigify/introduction.html}
\showURL{%
\tempurl}


\bibitem[Bolkart et~al\mbox{.}(2023)]%
        {bolkart2023instant}
\bibfield{author}{\bibinfo{person}{Timo Bolkart}, \bibinfo{person}{Tianye Li}, {and} \bibinfo{person}{Michael~J Black}.} \bibinfo{year}{2023}\natexlab{}.
\newblock \showarticletitle{Instant Multi-View Head Capture through Learnable Registration}. In \bibinfo{booktitle}{\emph{Proceedings of the IEEE/CVF Conference on Computer Vision and Pattern Recognition}}. \bibinfo{pages}{768--779}.
\newblock


\bibitem[Booth et~al\mbox{.}(2016)]%
        {booth20163d}
\bibfield{author}{\bibinfo{person}{James Booth}, \bibinfo{person}{Anastasios Roussos}, \bibinfo{person}{Stefanos Zafeiriou}, \bibinfo{person}{Allan Ponniah}, {and} \bibinfo{person}{David Dunaway}.} \bibinfo{year}{2016}\natexlab{}.
\newblock \showarticletitle{A 3d morphable model learnt from 10,000 faces}. In \bibinfo{booktitle}{\emph{Proceedings of the IEEE conference on computer vision and pattern recognition}}. \bibinfo{pages}{5543--5552}.
\newblock


\bibitem[Bradley et~al\mbox{.}(2010)]%
        {bradley2010high}
\bibfield{author}{\bibinfo{person}{Derek Bradley}, \bibinfo{person}{Wolfgang Heidrich}, \bibinfo{person}{Tiberiu Popa}, {and} \bibinfo{person}{Alla Sheffer}.} \bibinfo{year}{2010}\natexlab{}.
\newblock \showarticletitle{High resolution passive facial performance capture}.
\newblock In \bibinfo{booktitle}{\emph{ACM SIGGRAPH 2010 papers}}. \bibinfo{pages}{1--10}.
\newblock


\bibitem[Bulat and Tzimiropoulos(2017)]%
        {bulat2017far}
\bibfield{author}{\bibinfo{person}{Adrian Bulat} {and} \bibinfo{person}{Georgios Tzimiropoulos}.} \bibinfo{year}{2017}\natexlab{}.
\newblock \showarticletitle{How far are we from solving the 2d \& 3d face alignment problem? (and a dataset of 230,000 3d facial landmarks)}. In \bibinfo{booktitle}{\emph{Proceedings of the IEEE International Conference on Computer Vision}}. \bibinfo{pages}{1021--1030}.
\newblock


\bibitem[Cao et~al\mbox{.}(2022)]%
        {cao2022authentic}
\bibfield{author}{\bibinfo{person}{Chen Cao}, \bibinfo{person}{Tomas Simon}, \bibinfo{person}{Jin~Kyu Kim}, \bibinfo{person}{Gabe Schwartz}, \bibinfo{person}{Michael Zollhoefer}, \bibinfo{person}{Shun-Suke Saito}, \bibinfo{person}{Stephen Lombardi}, \bibinfo{person}{Shih-En Wei}, \bibinfo{person}{Danielle Belko}, \bibinfo{person}{Shoou-I Yu}, {et~al\mbox{.}}} \bibinfo{year}{2022}\natexlab{}.
\newblock \showarticletitle{Authentic volumetric avatars from a phone scan}.
\newblock \bibinfo{journal}{\emph{ACM Transactions on Graphics (TOG)}} \bibinfo{volume}{41}, \bibinfo{number}{4} (\bibinfo{year}{2022}), \bibinfo{pages}{1--19}.
\newblock


\bibitem[Chandran et~al\mbox{.}(2022)]%
        {chandran2022facial}
\bibfield{author}{\bibinfo{person}{Prashanth Chandran}, \bibinfo{person}{Gaspard Zoss}, \bibinfo{person}{Markus Gross}, \bibinfo{person}{Paulo Gotardo}, {and} \bibinfo{person}{Derek Bradley}.} \bibinfo{year}{2022}\natexlab{}.
\newblock \showarticletitle{Facial Animation with Disentangled Identity and Motion using Transformers}. In \bibinfo{booktitle}{\emph{ACM/Eurographics Symposium on Computer Animation 2022}}.
\newblock


\bibitem[Chaudhuri et~al\mbox{.}(2020)]%
        {chaudhuri2020personalized}
\bibfield{author}{\bibinfo{person}{Bindita Chaudhuri}, \bibinfo{person}{Noranart Vesdapunt}, \bibinfo{person}{Linda Shapiro}, {and} \bibinfo{person}{Baoyuan Wang}.} \bibinfo{year}{2020}\natexlab{}.
\newblock \showarticletitle{Personalized Face Modeling for Improved Face Reconstruction and Motion Retargeting}. In \bibinfo{booktitle}{\emph{Computer Vision -- ECCV 2020}}, \bibfield{editor}{\bibinfo{person}{Andrea Vedaldi}, \bibinfo{person}{Horst Bischof}, \bibinfo{person}{Thomas Brox}, {and} \bibinfo{person}{Jan-Michael Frahm}} (Eds.). \bibinfo{publisher}{Springer International Publishing}, \bibinfo{address}{Cham}, \bibinfo{pages}{142--160}.
\newblock


\bibitem[Chrysos et~al\mbox{.}(2018)]%
        {chrysos2018comprehensive}
\bibfield{author}{\bibinfo{person}{Grigorios~G Chrysos}, \bibinfo{person}{Epameinondas Antonakos}, \bibinfo{person}{Patrick Snape}, \bibinfo{person}{Akshay Asthana}, {and} \bibinfo{person}{Stefanos Zafeiriou}.} \bibinfo{year}{2018}\natexlab{}.
\newblock \showarticletitle{A comprehensive performance evaluation of deformable face tracking “in-the-wild”}.
\newblock \bibinfo{journal}{\emph{International Journal of Computer Vision}} \bibinfo{volume}{126}, \bibinfo{number}{2} (\bibinfo{year}{2018}), \bibinfo{pages}{198--232}.
\newblock


\bibitem[Cong et~al\mbox{.}(2015)]%
        {cong2015fully}
\bibfield{author}{\bibinfo{person}{Matthew Cong}, \bibinfo{person}{Michael Bao}, \bibinfo{person}{Jane~L E}, \bibinfo{person}{Kiran~S Bhat}, {and} \bibinfo{person}{Ronald Fedkiw}.} \bibinfo{year}{2015}\natexlab{}.
\newblock \showarticletitle{Fully automatic generation of anatomical face simulation models}. In \bibinfo{booktitle}{\emph{Proceedings of the 14th ACM SIGGRAPH/Eurographics Symposium on Computer Animation}}. \bibinfo{pages}{175--183}.
\newblock


\bibitem[{Daz3D}(2023)]%
        {Daz3D}
\bibfield{author}{\bibinfo{person}{{Daz3D}}.} \bibinfo{year}{2023}\natexlab{}.
\newblock \bibinfo{booktitle}{\emph{FaceGen}}.
\newblock
\urldef\tempurl%
\url{https://www.daz3d.com/easy-character-creation-with-facegen}
\showURL{%
\tempurl}


\bibitem[Debevec(2012)]%
        {debevec2012light}
\bibfield{author}{\bibinfo{person}{Paul Debevec}.} \bibinfo{year}{2012}\natexlab{}.
\newblock \showarticletitle{The light stages and their applications to photoreal digital actors}.
\newblock \bibinfo{journal}{\emph{SIGGRAPH Asia Technical Briefs}} \bibinfo{volume}{2}, \bibinfo{number}{4} (\bibinfo{year}{2012}), \bibinfo{pages}{1--6}.
\newblock


\bibitem[Debevec et~al\mbox{.}(2000)]%
        {debevec2000acquiring}
\bibfield{author}{\bibinfo{person}{Paul Debevec}, \bibinfo{person}{Tim Hawkins}, \bibinfo{person}{Chris Tchou}, \bibinfo{person}{Haarm-Pieter Duiker}, \bibinfo{person}{Westley Sarokin}, {and} \bibinfo{person}{Mark Sagar}.} \bibinfo{year}{2000}\natexlab{}.
\newblock \showarticletitle{Acquiring the reflectance field of a human face}. In \bibinfo{booktitle}{\emph{Proceedings of the 27th annual conference on Computer graphics and interactive techniques}}. \bibinfo{pages}{145--156}.
\newblock


\bibitem[Dib et~al\mbox{.}(2023)]%
        {dib2023s2f2}
\bibfield{author}{\bibinfo{person}{Abdallah Dib}, \bibinfo{person}{Junghyun Ahn}, \bibinfo{person}{Cedric Thebault}, \bibinfo{person}{Philippe-Henri Gosselin}, {and} \bibinfo{person}{Louis Chevallier}.} \bibinfo{year}{2023}\natexlab{}.
\newblock \showarticletitle{S2f2: self-supervised high fidelity face reconstruction from monocular image}. In \bibinfo{booktitle}{\emph{2023 IEEE 17th International Conference on Automatic Face and Gesture Recognition (FG)}}. IEEE, \bibinfo{pages}{1--8}.
\newblock


\bibitem[Doi and Koide(1991)]%
        {doi1991efficient}
\bibfield{author}{\bibinfo{person}{Akio Doi} {and} \bibinfo{person}{Akio Koide}.} \bibinfo{year}{1991}\natexlab{}.
\newblock \showarticletitle{An efficient method of triangulating equi-valued surfaces by using tetrahedral cells}.
\newblock \bibinfo{journal}{\emph{IEICE TRANSACTIONS on Information and Systems}} \bibinfo{volume}{74}, \bibinfo{number}{1} (\bibinfo{year}{1991}), \bibinfo{pages}{214--224}.
\newblock


\bibitem[Dou and Kakadiaris(2018)]%
        {dou2018multi}
\bibfield{author}{\bibinfo{person}{Pengfei Dou} {and} \bibinfo{person}{Ioannis~A Kakadiaris}.} \bibinfo{year}{2018}\natexlab{}.
\newblock \showarticletitle{Multi-view 3D face reconstruction with deep recurrent neural networks}.
\newblock \bibinfo{journal}{\emph{Image and Vision Computing}}  \bibinfo{volume}{80} (\bibinfo{year}{2018}), \bibinfo{pages}{80--91}.
\newblock


\bibitem[Dou et~al\mbox{.}(2017)]%
        {dou2017endtoend}
\bibfield{author}{\bibinfo{person}{P. Dou}, \bibinfo{person}{S.~K. Shah}, {and} \bibinfo{person}{I.~A. Kakadiaris}.} \bibinfo{year}{2017}\natexlab{}.
\newblock \showarticletitle{End-to-End 3D Face Reconstruction with Deep Neural Networks}. In \bibinfo{booktitle}{\emph{2017 IEEE Conference on Computer Vision and Pattern Recognition (CVPR)}}. \bibinfo{publisher}{IEEE Computer Society}, \bibinfo{address}{Los Alamitos, CA, USA}, \bibinfo{pages}{1503--1512}.
\newblock
\showISSN{1063-6919}


\bibitem[{Epic Games}(2023)]%
        {unrealengine}
\bibfield{author}{\bibinfo{person}{{Epic Games}}.} \bibinfo{year}{2023}\natexlab{}.
\newblock \bibinfo{booktitle}{\emph{Unreal Engine}}.
\newblock
\urldef\tempurl%
\url{https://www.unrealengine.com}
\showURL{%
\tempurl}


\bibitem[Feng et~al\mbox{.}(2022)]%
        {feng2022trust}
\bibfield{author}{\bibinfo{person}{Haiwen Feng}, \bibinfo{person}{Timo Bolkart}, \bibinfo{person}{Joachim Tesch}, \bibinfo{person}{Michael~J. Black}, {and} \bibinfo{person}{Victoria Abrevaya}.} \bibinfo{year}{2022}\natexlab{}.
\newblock \showarticletitle{Towards Racially Unbiased Skin Tone Estimation via Scene Disambiguation}. In \bibinfo{booktitle}{\emph{European Conference on Computer Vision}}.
\newblock


\bibitem[Field(1988)]%
        {field1988laplacian}
\bibfield{author}{\bibinfo{person}{David~A Field}.} \bibinfo{year}{1988}\natexlab{}.
\newblock \showarticletitle{Laplacian smoothing and Delaunay triangulations}.
\newblock \bibinfo{journal}{\emph{Communications in applied numerical methods}} \bibinfo{volume}{4}, \bibinfo{number}{6} (\bibinfo{year}{1988}), \bibinfo{pages}{709--712}.
\newblock


\bibitem[Franc(2023)]%
        {riglogic22}
\bibfield{author}{\bibinfo{person}{Andrean Franc}.} \bibinfo{year}{2023}\natexlab{}.
\newblock \bibinfo{title}{RIG LOGIC: RUNTIME EVALUATION OF METAHUMAN FACE RIGS}.
\newblock \bibinfo{howpublished}{\url{https://cdn2.unrealengine.com/rig-logic-whitepaper-v2-5c9f23f7e210.pdf}}.
\newblock
\newblock
\shownote{Accessed: 2023-05-30}.


\bibitem[Gafni et~al\mbox{.}(2021)]%
        {gafni2021nerface}
\bibfield{author}{\bibinfo{person}{Guy Gafni}, \bibinfo{person}{Justus Thies}, \bibinfo{person}{Michael Zollh{\"o}fer}, {and} \bibinfo{person}{Matthias Nie{\ss}ner}.} \bibinfo{year}{2021}\natexlab{}.
\newblock \showarticletitle{Dynamic Neural Radiance Fields for Monocular 4D Facial Avatar Reconstruction}. In \bibinfo{booktitle}{\emph{Proceedings of the IEEE/CVF Conference on Computer Vision and Pattern Recognition (CVPR)}}. \bibinfo{pages}{8649--8658}.
\newblock


\bibitem[Gao et~al\mbox{.}(2022)]%
        {gao2022reconstructing}
\bibfield{author}{\bibinfo{person}{Xuan Gao}, \bibinfo{person}{Chenglai Zhong}, \bibinfo{person}{Jun Xiang}, \bibinfo{person}{Yang Hong}, \bibinfo{person}{Yudong Guo}, {and} \bibinfo{person}{Juyong Zhang}.} \bibinfo{year}{2022}\natexlab{}.
\newblock \showarticletitle{Reconstructing personalized semantic facial nerf models from monocular video}.
\newblock \bibinfo{journal}{\emph{ACM Transactions on Graphics (TOG)}} \bibinfo{volume}{41}, \bibinfo{number}{6} (\bibinfo{year}{2022}), \bibinfo{pages}{1--12}.
\newblock


\bibitem[Garg et~al\mbox{.}(2013)]%
        {garg2013dense}
\bibfield{author}{\bibinfo{person}{Ravi Garg}, \bibinfo{person}{Anastasios Roussos}, {and} \bibinfo{person}{Lourdes Agapito}.} \bibinfo{year}{2013}\natexlab{}.
\newblock \showarticletitle{Dense variational reconstruction of non-rigid surfaces from monocular video}. In \bibinfo{booktitle}{\emph{Proceedings of the IEEE Conference on computer vision and pattern recognition}}. \bibinfo{pages}{1272--1279}.
\newblock


\bibitem[Garrido et~al\mbox{.}(2016)]%
        {garrido2016reconstruction}
\bibfield{author}{\bibinfo{person}{Pablo Garrido}, \bibinfo{person}{Michael Zollh{\"o}fer}, \bibinfo{person}{Dan Casas}, \bibinfo{person}{Levi Valgaerts}, \bibinfo{person}{Kiran Varanasi}, \bibinfo{person}{Patrick P{\'e}rez}, {and} \bibinfo{person}{Christian Theobalt}.} \bibinfo{year}{2016}\natexlab{}.
\newblock \showarticletitle{Reconstruction of personalized 3D face rigs from monocular video}.
\newblock \bibinfo{journal}{\emph{ACM Transactions on Graphics (TOG)}} \bibinfo{volume}{35}, \bibinfo{number}{3} (\bibinfo{year}{2016}), \bibinfo{pages}{1--15}.
\newblock


\bibitem[Ghosh et~al\mbox{.}(2011)]%
        {ghosh2011multiview}
\bibfield{author}{\bibinfo{person}{Abhijeet Ghosh}, \bibinfo{person}{Graham Fyffe}, \bibinfo{person}{Borom Tunwattanapong}, \bibinfo{person}{Jay Busch}, \bibinfo{person}{Xueming Yu}, {and} \bibinfo{person}{Paul Debevec}.} \bibinfo{year}{2011}\natexlab{}.
\newblock \showarticletitle{Multiview face capture using polarized spherical gradient illumination}. In \bibinfo{booktitle}{\emph{Proceedings of the 2011 SIGGRAPH Asia Conference}}. \bibinfo{pages}{1--10}.
\newblock


\bibitem[Gower(1975)]%
        {gower1975generalized}
\bibfield{author}{\bibinfo{person}{John~C Gower}.} \bibinfo{year}{1975}\natexlab{}.
\newblock \showarticletitle{Generalized procrustes analysis}.
\newblock \bibinfo{journal}{\emph{Psychometrika}}  \bibinfo{volume}{40} (\bibinfo{year}{1975}), \bibinfo{pages}{33--51}.
\newblock


\bibitem[Guo et~al\mbox{.}(2021)]%
        {guo2021adnerf}
\bibfield{author}{\bibinfo{person}{Yudong Guo}, \bibinfo{person}{Keyu Chen}, \bibinfo{person}{Sen Liang}, \bibinfo{person}{Yongjin Liu}, \bibinfo{person}{Hujun Bao}, {and} \bibinfo{person}{Juyong Zhang}.} \bibinfo{year}{2021}\natexlab{}.
\newblock \showarticletitle{AD-NeRF: Audio Driven Neural Radiance Fields for Talking Head Synthesis}. In \bibinfo{booktitle}{\emph{{IEEE/CVF} International Conference on Computer Vision (ICCV)}}.
\newblock


\bibitem[Han et~al\mbox{.}(2023)]%
        {han2023learning}
\bibfield{author}{\bibinfo{person}{Yuxuan Han}, \bibinfo{person}{Zhibo Wang}, {and} \bibinfo{person}{Feng Xu}.} \bibinfo{year}{2023}\natexlab{}.
\newblock \showarticletitle{Learning a 3D Morphable Face Reflectance Model From Low-Cost Data}. In \bibinfo{booktitle}{\emph{Proceedings of the IEEE/CVF Conference on Computer Vision and Pattern Recognition}}. \bibinfo{pages}{8598--8608}.
\newblock


\bibitem[{HDReye Technologies Inc.}(2023)]%
        {hdreye}
\bibfield{author}{\bibinfo{person}{{HDReye Technologies Inc.}}} \bibinfo{year}{2023}\natexlab{}.
\newblock \bibinfo{booktitle}{\emph{HDReye}}.
\newblock
\urldef\tempurl%
\url{https://hdreye.app/}
\showURL{%
\tempurl}


\bibitem[Hendler et~al\mbox{.}(2018)]%
        {hendler2018avengers}
\bibfield{author}{\bibinfo{person}{Darren Hendler}, \bibinfo{person}{Lucio Moser}, \bibinfo{person}{Rishabh Battulwar}, \bibinfo{person}{David Corral}, \bibinfo{person}{Phil Cramer}, \bibinfo{person}{Ron Miller}, \bibinfo{person}{Rickey Cloudsdale}, {and} \bibinfo{person}{Doug Roble}.} \bibinfo{year}{2018}\natexlab{}.
\newblock \showarticletitle{Avengers: Capturing Thanos's Complex Face}. In \bibinfo{booktitle}{\emph{ACM SIGGRAPH 2018 Talks}} (Vancouver, British Columbia, Canada) \emph{(\bibinfo{series}{SIGGRAPH '18})}. \bibinfo{publisher}{Association for Computing Machinery}, \bibinfo{address}{New York, NY, USA}, Article \bibinfo{articleno}{58}, \bibinfo{numpages}{2}~pages.
\newblock
\showISBNx{9781450358200}
\urldef\tempurl%
\url{https://doi.org/10.1145/3214745.3214766}
\showDOI{\tempurl}


\bibitem[Holz et~al\mbox{.}(2015)]%
        {holz2015registration}
\bibfield{author}{\bibinfo{person}{Dirk Holz}, \bibinfo{person}{Alexandru~E Ichim}, \bibinfo{person}{Federico Tombari}, \bibinfo{person}{Radu~B Rusu}, {and} \bibinfo{person}{Sven Behnke}.} \bibinfo{year}{2015}\natexlab{}.
\newblock \showarticletitle{Registration with the point cloud library: A modular framework for aligning in 3-D}.
\newblock \bibinfo{journal}{\emph{IEEE Robotics \& Automation Magazine}} \bibinfo{volume}{22}, \bibinfo{number}{4} (\bibinfo{year}{2015}), \bibinfo{pages}{110--124}.
\newblock


\bibitem[Hyprsense(2021)]%
        {hyprsense2020}
Hyprsense \bibinfo{year}{2021}\natexlab{}.
\newblock \bibinfo{booktitle}{\emph{Hyprface Landmark Tracker}}.
\newblock
\urldef\tempurl%
\url{https://medium.com/@hyprsense}
\showURL{%
Retrieved Dec 23, 2023 from \tempurl}


\bibitem[Ichim et~al\mbox{.}(2015)]%
        {ichim2015dynamic}
\bibfield{author}{\bibinfo{person}{Alexandru~Eugen Ichim}, \bibinfo{person}{Sofien Bouaziz}, {and} \bibinfo{person}{Mark Pauly}.} \bibinfo{year}{2015}\natexlab{}.
\newblock \showarticletitle{Dynamic 3D avatar creation from hand-held video input}.
\newblock \bibinfo{journal}{\emph{ACM Transactions on Graphics (ToG)}} \bibinfo{volume}{34}, \bibinfo{number}{4} (\bibinfo{year}{2015}), \bibinfo{pages}{1--14}.
\newblock


\bibitem[{Insta360}(2023)]%
        {insta360x3}
\bibfield{author}{\bibinfo{person}{{Insta360}}.} \bibinfo{year}{2023}\natexlab{}.
\newblock \bibinfo{booktitle}{\emph{insta360 X3}}.
\newblock
\urldef\tempurl%
\url{https://store.insta360.com/product/x3}
\showURL{%
\tempurl}


\bibitem[Jackson et~al\mbox{.}(2017)]%
        {jackson2017vrn}
\bibfield{author}{\bibinfo{person}{Aaron~S Jackson}, \bibinfo{person}{Adrian Bulat}, \bibinfo{person}{Vasileios Argyriou}, {and} \bibinfo{person}{Georgios Tzimiropoulos}.} \bibinfo{year}{2017}\natexlab{}.
\newblock \showarticletitle{Large Pose 3D Face Reconstruction from a Single Image via Direct Volumetric CNN Regression}.
\newblock \bibinfo{journal}{\emph{International Conference on Computer Vision}} (\bibinfo{year}{2017}).
\newblock


\bibitem[Joo et~al\mbox{.}(2017)]%
        {joo2017panoptic}
\bibfield{author}{\bibinfo{person}{Hanbyul Joo}, \bibinfo{person}{Tomas Simon}, \bibinfo{person}{Xulong Li}, \bibinfo{person}{Hao Liu}, \bibinfo{person}{Lei Tan}, \bibinfo{person}{Lin Gui}, \bibinfo{person}{Sean Banerjee}, \bibinfo{person}{Timothy~Scott Godisart}, \bibinfo{person}{Bart Nabbe}, \bibinfo{person}{Iain Matthews}, \bibinfo{person}{Takeo Kanade}, \bibinfo{person}{Shohei Nobuhara}, {and} \bibinfo{person}{Yaser Sheikh}.} \bibinfo{year}{2017}\natexlab{}.
\newblock \showarticletitle{Panoptic Studio: A Massively Multiview System for Social Interaction Capture}.
\newblock \bibinfo{journal}{\emph{IEEE Transactions on Pattern Analysis and Machine Intelligence}} (\bibinfo{year}{2017}).
\newblock


\bibitem[Kazhdan and Hoppe(2013)]%
        {kazhdan2013screened}
\bibfield{author}{\bibinfo{person}{Michael Kazhdan} {and} \bibinfo{person}{Hugues Hoppe}.} \bibinfo{year}{2013}\natexlab{}.
\newblock \showarticletitle{Screened poisson surface reconstruction}.
\newblock \bibinfo{journal}{\emph{ACM Transactions on Graphics (ToG)}} \bibinfo{volume}{32}, \bibinfo{number}{3} (\bibinfo{year}{2013}), \bibinfo{pages}{1--13}.
\newblock


\bibitem[Kemelmacher-Shlizerman(2013)]%
        {kemelmacher2013internet}
\bibfield{author}{\bibinfo{person}{Ira Kemelmacher-Shlizerman}.} \bibinfo{year}{2013}\natexlab{}.
\newblock \showarticletitle{Internet based morphable model}. In \bibinfo{booktitle}{\emph{Proceedings of the IEEE international conference on computer vision}}. \bibinfo{pages}{3256--3263}.
\newblock


\bibitem[Kemelmacher-Shlizerman and Seitz(2011)]%
        {kemelmacher2011face}
\bibfield{author}{\bibinfo{person}{Ira Kemelmacher-Shlizerman} {and} \bibinfo{person}{Steven~M Seitz}.} \bibinfo{year}{2011}\natexlab{}.
\newblock \showarticletitle{Face reconstruction in the wild}. In \bibinfo{booktitle}{\emph{2011 international conference on computer vision}}. IEEE, \bibinfo{pages}{1746--1753}.
\newblock


\bibitem[Kim et~al\mbox{.}(2021)]%
        {kim2021learning}
\bibfield{author}{\bibinfo{person}{Jongyoo Kim}, \bibinfo{person}{Jiaolong Yang}, {and} \bibinfo{person}{Xin Tong}.} \bibinfo{year}{2021}\natexlab{}.
\newblock \showarticletitle{Learning high-fidelity face texture completion without complete face texture}. In \bibinfo{booktitle}{\emph{Proceedings of the IEEE/CVF International Conference on Computer Vision}}. \bibinfo{pages}{13990--13999}.
\newblock


\bibitem[Konolige(1998)]%
        {konolige1998small}
\bibfield{author}{\bibinfo{person}{Kurt Konolige}.} \bibinfo{year}{1998}\natexlab{}.
\newblock \showarticletitle{Small vision systems: Hardware and implementation}. In \bibinfo{booktitle}{\emph{Robotics Research: The Eighth International Symposium}}. Springer, \bibinfo{pages}{203--212}.
\newblock


\bibitem[Lattas et~al\mbox{.}(2022)]%
        {lattas2022practical}
\bibfield{author}{\bibinfo{person}{Alexandros Lattas}, \bibinfo{person}{Yiming Lin}, \bibinfo{person}{Jayanth Kannan}, \bibinfo{person}{Ekin Ozturk}, \bibinfo{person}{Luca Filipi}, \bibinfo{person}{Giuseppe~Claudio Guarnera}, \bibinfo{person}{Gaurav Chawla}, {and} \bibinfo{person}{Abhijeet Ghosh}.} \bibinfo{year}{2022}\natexlab{}.
\newblock \showarticletitle{Practical and Scalable Desktop-Based High-Quality Facial Capture}. In \bibinfo{booktitle}{\emph{Computer Vision -- ECCV 2022}}. \bibinfo{publisher}{Springer Nature Switzerland}, \bibinfo{address}{Cham}, \bibinfo{pages}{522--537}.
\newblock


\bibitem[Lewis et~al\mbox{.}(2014)]%
        {lewis2014practice}
\bibfield{author}{\bibinfo{person}{John~P Lewis}, \bibinfo{person}{Ken Anjyo}, \bibinfo{person}{Taehyun Rhee}, \bibinfo{person}{Mengjie Zhang}, \bibinfo{person}{Frederic~H Pighin}, {and} \bibinfo{person}{Zhigang Deng}.} \bibinfo{year}{2014}\natexlab{}.
\newblock \showarticletitle{Practice and theory of blendshape facial models.}
\newblock \bibinfo{journal}{\emph{Eurographics (State of the Art Reports)}} \bibinfo{volume}{1}, \bibinfo{number}{8} (\bibinfo{year}{2014}), \bibinfo{pages}{2}.
\newblock


\bibitem[Li et~al\mbox{.}(2022)]%
        {li2022implicit}
\bibfield{author}{\bibinfo{person}{Moran Li}, \bibinfo{person}{Haibin Huang}, \bibinfo{person}{Yi Zheng}, \bibinfo{person}{Mengtian Li}, \bibinfo{person}{Nong Sang}, {and} \bibinfo{person}{Chongyang Ma}.} \bibinfo{year}{2022}\natexlab{}.
\newblock \showarticletitle{Implicit Neural Deformation for Sparse-View Face Reconstruction}.
\newblock \bibinfo{journal}{\emph{Computer Graphics Forum}} \bibinfo{volume}{41}, \bibinfo{number}{7} (\bibinfo{year}{2022}), \bibinfo{pages}{601--610}.
\newblock
\urldef\tempurl%
\url{https://doi.org/10.1111/cgf.14704}
\showDOI{\tempurl}


\bibitem[Li et~al\mbox{.}(2017)]%
        {li2017learning}
\bibfield{author}{\bibinfo{person}{Tianye Li}, \bibinfo{person}{Timo Bolkart}, \bibinfo{person}{Michael~J Black}, \bibinfo{person}{Hao Li}, {and} \bibinfo{person}{Javier Romero}.} \bibinfo{year}{2017}\natexlab{}.
\newblock \showarticletitle{Learning a model of facial shape and expression from 4D scans.}
\newblock \bibinfo{journal}{\emph{ACM Trans. Graph.}} \bibinfo{volume}{36}, \bibinfo{number}{6} (\bibinfo{year}{2017}), \bibinfo{pages}{194--1}.
\newblock


\bibitem[Liang et~al\mbox{.}(2016)]%
        {liang2016head}
\bibfield{author}{\bibinfo{person}{Shu Liang}, \bibinfo{person}{Linda~G Shapiro}, {and} \bibinfo{person}{Ira Kemelmacher-Shlizerman}.} \bibinfo{year}{2016}\natexlab{}.
\newblock \showarticletitle{Head reconstruction from internet photos}. In \bibinfo{booktitle}{\emph{Computer Vision--ECCV 2016: 14th European Conference, Amsterdam, The Netherlands, October 11-14, 2016, Proceedings, Part II 14}}. Springer, \bibinfo{pages}{360--374}.
\newblock


\bibitem[Lin et~al\mbox{.}(2023)]%
        {lin2023single}
\bibfield{author}{\bibinfo{person}{Connor Lin}, \bibinfo{person}{Koki Nagano}, \bibinfo{person}{Jan Kautz}, \bibinfo{person}{Eric Chan}, \bibinfo{person}{Umar Iqbal}, \bibinfo{person}{Leonidas Guibas}, \bibinfo{person}{Gordon Wetzstein}, {and} \bibinfo{person}{Sameh Khamis}.} \bibinfo{year}{2023}\natexlab{}.
\newblock \showarticletitle{Single-shot implicit morphable faces with consistent texture parameterization}. In \bibinfo{booktitle}{\emph{ACM SIGGRAPH 2023 Conference Proceedings}}. \bibinfo{pages}{1--12}.
\newblock


\bibitem[Lin et~al\mbox{.}(2021)]%
        {lin2021meingame}
\bibfield{author}{\bibinfo{person}{Jiangke Lin}, \bibinfo{person}{Yi Yuan}, {and} \bibinfo{person}{Zhengxia Zou}.} \bibinfo{year}{2021}\natexlab{}.
\newblock \showarticletitle{Meingame: Create a game character face from a single portrait}. In \bibinfo{booktitle}{\emph{Proceedings of the AAAI Conference on Artificial Intelligence}}, Vol.~\bibinfo{volume}{35}. \bibinfo{pages}{311--319}.
\newblock


\bibitem[Lin et~al\mbox{.}(2022)]%
        {lin2022leveraging}
\bibfield{author}{\bibinfo{person}{Winnie Lin}, \bibinfo{person}{Yilin Zhu}, \bibinfo{person}{Demi Guo}, {and} \bibinfo{person}{Ron Fedkiw}.} \bibinfo{year}{2022}\natexlab{}.
\newblock \bibinfo{title}{Leveraging Deepfakes to Close the Domain Gap between Real and Synthetic Images in Facial Capture Pipelines}.
\newblock
\newblock
\showeprint[arxiv]{2204.10746}~[cs.CV]


\bibitem[Liu et~al\mbox{.}(2022)]%
        {liu2022rapid}
\bibfield{author}{\bibinfo{person}{Shichen Liu}, \bibinfo{person}{Yunxuan Cai}, \bibinfo{person}{Haiwei Chen}, \bibinfo{person}{Yichao Zhou}, {and} \bibinfo{person}{Yajie Zhao}.} \bibinfo{year}{2022}\natexlab{}.
\newblock \showarticletitle{Rapid Face Asset Acquisition with Recurrent Feature Alignment}.
\newblock \bibinfo{journal}{\emph{ACM Trans. Graph.}} \bibinfo{volume}{41}, \bibinfo{number}{6}, Article \bibinfo{articleno}{214} (\bibinfo{date}{nov} \bibinfo{year}{2022}), \bibinfo{numpages}{17}~pages.
\newblock
\showISSN{0730-0301}


\bibitem[M2M(2023)]%
        {m2m}
M2M \bibinfo{year}{2023}\natexlab{}.
\newblock \bibinfo{booktitle}{\emph{Mesh to Metahuman for Unreal Engine}}.
\newblock
\urldef\tempurl%
\url{https://dev.epicgames.com/documentation/en-us/metahuman/mesh-to-metahuman-for-unreal-engine}
\showURL{%
Retrieved Dec 23, 2023 from \tempurl}


\bibitem[MHC(2023)]%
        {mhc}
MHC \bibinfo{year}{2023}\natexlab{}.
\newblock \bibinfo{booktitle}{\emph{Metahuman}}.
\newblock
\urldef\tempurl%
\url{https://dev.epicgames.com/documentation/en-US/metahuman/metahuman-documentation}
\showURL{%
Retrieved Dec 23, 2023 from \tempurl}


\bibitem[Mildenhall et~al\mbox{.}(2021)]%
        {mildenhall2021nerf}
\bibfield{author}{\bibinfo{person}{Ben Mildenhall}, \bibinfo{person}{Pratul~P Srinivasan}, \bibinfo{person}{Matthew Tancik}, \bibinfo{person}{Jonathan~T Barron}, \bibinfo{person}{Ravi Ramamoorthi}, {and} \bibinfo{person}{Ren Ng}.} \bibinfo{year}{2021}\natexlab{}.
\newblock \showarticletitle{Nerf: Representing scenes as neural radiance fields for view synthesis}.
\newblock \bibinfo{journal}{\emph{Commun. ACM}} \bibinfo{volume}{65}, \bibinfo{number}{1} (\bibinfo{year}{2021}), \bibinfo{pages}{99--106}.
\newblock


\bibitem[Piotraschke and Blanz(2016)]%
        {piotraschke2016automated}
\bibfield{author}{\bibinfo{person}{Marcel Piotraschke} {and} \bibinfo{person}{Volker Blanz}.} \bibinfo{year}{2016}\natexlab{}.
\newblock \showarticletitle{Automated 3d face reconstruction from multiple images using quality measures}. In \bibinfo{booktitle}{\emph{Proceedings of the IEEE conference on computer vision and pattern recognition}}. \bibinfo{pages}{3418--3427}.
\newblock


\bibitem[Qin et~al\mbox{.}(2023)]%
        {10.1145/3588432.3591556}
\bibfield{author}{\bibinfo{person}{Dafei Qin}, \bibinfo{person}{Jun Saito}, \bibinfo{person}{Noam Aigerman}, \bibinfo{person}{Thibault Groueix}, {and} \bibinfo{person}{Taku Komura}.} \bibinfo{year}{2023}\natexlab{}.
\newblock \showarticletitle{Neural Face Rigging for Animating and Retargeting Facial Meshes in the Wild}. In \bibinfo{booktitle}{\emph{ACM SIGGRAPH 2023 Conference Proceedings}} (Los Angeles, CA, USA) \emph{(\bibinfo{series}{SIGGRAPH '23})}. \bibinfo{publisher}{Association for Computing Machinery}, \bibinfo{address}{New York, NY, USA}, Article \bibinfo{articleno}{68}, \bibinfo{numpages}{11}~pages.
\newblock
\showISBNx{9798400701597}
\urldef\tempurl%
\url{https://doi.org/10.1145/3588432.3591556}
\showDOI{\tempurl}


\bibitem[Rainer et~al\mbox{.}(2023)]%
        {rainer2023neural}
\bibfield{author}{\bibinfo{person}{Gilles Rainer}, \bibinfo{person}{Lewis Bridgeman}, {and} \bibinfo{person}{Abhijeet Ghosh}.} \bibinfo{year}{2023}\natexlab{}.
\newblock \showarticletitle{Neural shading fields for efficient facial inverse rendering}. In \bibinfo{booktitle}{\emph{Computer Graphics Forum}}, Vol.~\bibinfo{volume}{42}. Wiley Online Library, \bibinfo{pages}{e14943}.
\newblock


\bibitem[Ramamoorthi and Hanrahan(2001)]%
        {ramamoorthi2001efficient}
\bibfield{author}{\bibinfo{person}{Ravi Ramamoorthi} {and} \bibinfo{person}{Pat Hanrahan}.} \bibinfo{year}{2001}\natexlab{}.
\newblock \showarticletitle{An Efficient Representation for Irradiance Environment Maps}. In \bibinfo{booktitle}{\emph{Proceedings of the 28th Annual Conference on Computer Graphics and Interactive Techniques}} \emph{(\bibinfo{series}{SIGGRAPH '01})}. \bibinfo{publisher}{Association for Computing Machinery}, \bibinfo{address}{New York, NY, USA}, \bibinfo{pages}{497–500}.
\newblock
\showISBNx{158113374X}


\bibitem[Rasmussen et~al\mbox{.}(2003)]%
        {rasmussen2003smoke}
\bibfield{author}{\bibinfo{person}{Nick Rasmussen}, \bibinfo{person}{Duc~Quang Nguyen}, \bibinfo{person}{Willi Geiger}, {and} \bibinfo{person}{Ronald Fedkiw}.} \bibinfo{year}{2003}\natexlab{}.
\newblock \showarticletitle{Smoke simulation for large scale phenomena}.
\newblock In \bibinfo{booktitle}{\emph{ACM SIGGRAPH 2003 Papers}}. \bibinfo{pages}{703--707}.
\newblock


\bibitem[Ravi et~al\mbox{.}(2020)]%
        {ravi2020pytorch3d}
\bibfield{author}{\bibinfo{person}{Nikhila Ravi}, \bibinfo{person}{Jeremy Reizenstein}, \bibinfo{person}{David Novotny}, \bibinfo{person}{Taylor Gordon}, \bibinfo{person}{Wan-Yen Lo}, \bibinfo{person}{Justin Johnson}, {and} \bibinfo{person}{Georgia Gkioxari}.} \bibinfo{year}{2020}\natexlab{}.
\newblock \showarticletitle{Accelerating 3D Deep Learning with PyTorch3D}.
\newblock \bibinfo{journal}{\emph{arXiv:2007.08501}} (\bibinfo{year}{2020}).
\newblock


\bibitem[{Reallusion}(2023)]%
        {Headshot2}
\bibfield{author}{\bibinfo{person}{{Reallusion}}.} \bibinfo{year}{2023}\natexlab{}.
\newblock \bibinfo{booktitle}{\emph{Headshot2}}.
\newblock
\urldef\tempurl%
\url{https://www.reallusion.com/character-creator/headshot/}
\showURL{%
\tempurl}


\bibitem[Ren et~al\mbox{.}(2023)]%
        {ren2023improving}
\bibfield{author}{\bibinfo{person}{Xingyu Ren}, \bibinfo{person}{Jiankang Deng}, \bibinfo{person}{Chao Ma}, \bibinfo{person}{Yichao Yan}, {and} \bibinfo{person}{Xiaokang Yang}.} \bibinfo{year}{2023}\natexlab{}.
\newblock \showarticletitle{Improving Fairness in Facial Albedo Estimation via Visual-Textual Cues}. In \bibinfo{booktitle}{\emph{Proceedings of the IEEE/CVF Conference on Computer Vision and Pattern Recognition}}. \bibinfo{pages}{4511--4520}.
\newblock


\bibitem[Richardson et~al\mbox{.}(2016)]%
        {richardson20163d}
\bibfield{author}{\bibinfo{person}{E. Richardson}, \bibinfo{person}{M. Sela}, {and} \bibinfo{person}{R. Kimmel}.} \bibinfo{year}{2016}\natexlab{}.
\newblock \showarticletitle{3D Face Reconstruction by Learning from Synthetic Data}. In \bibinfo{booktitle}{\emph{2016 Fourth International Conference on 3D Vision (3DV)}}. \bibinfo{publisher}{IEEE Computer Society}, \bibinfo{address}{Los Alamitos, CA, USA}, \bibinfo{pages}{460--469}.
\newblock


\bibitem[{RICOH360}(2023)]%
        {ricohtheta}
\bibfield{author}{\bibinfo{person}{{RICOH360}}.} \bibinfo{year}{2023}\natexlab{}.
\newblock \bibinfo{booktitle}{\emph{RICOH THETA}}.
\newblock
\urldef\tempurl%
\url{https://www.ricoh360.com/theta/}
\showURL{%
\tempurl}


\bibitem[Riviere et~al\mbox{.}(2020)]%
        {riviere2020}
\bibfield{author}{\bibinfo{person}{J\'{e}r\'{e}my Riviere}, \bibinfo{person}{Paulo Gotardo}, \bibinfo{person}{Derek Bradley}, \bibinfo{person}{Abhijeet Ghosh}, {and} \bibinfo{person}{Thabo Beeler}.} \bibinfo{year}{2020}\natexlab{}.
\newblock \showarticletitle{Single-Shot High-Quality Facial Geometry and Skin Appearance Capture}.
\newblock \bibinfo{journal}{\emph{ACM Trans. Graph.}} \bibinfo{volume}{39}, \bibinfo{number}{4}, Article \bibinfo{articleno}{81} (\bibinfo{date}{jul} \bibinfo{year}{2020}), \bibinfo{numpages}{12}~pages.
\newblock
\showISSN{0730-0301}
\urldef\tempurl%
\url{https://doi.org/10.1145/3386569.3392464}
\showDOI{\tempurl}


\bibitem[Romdhani and Vetter(2003)]%
        {romdhani2003efficient}
\bibfield{author}{\bibinfo{person}{Romdhani} {and} \bibinfo{person}{Vetter}.} \bibinfo{year}{2003}\natexlab{}.
\newblock \showarticletitle{Efficient, robust and accurate fitting of a 3D morphable model}. In \bibinfo{booktitle}{\emph{Proceedings Ninth IEEE International Conference on Computer Vision}}. \bibinfo{pages}{59--66 vol.1}.
\newblock
\urldef\tempurl%
\url{https://doi.org/10.1109/ICCV.2003.1238314}
\showDOI{\tempurl}


\bibitem[Roth et~al\mbox{.}(2016)]%
        {roth2016adaptive}
\bibfield{author}{\bibinfo{person}{Joseph Roth}, \bibinfo{person}{Yiying Tong}, {and} \bibinfo{person}{Xiaoming Liu}.} \bibinfo{year}{2016}\natexlab{}.
\newblock \showarticletitle{Adaptive 3D face reconstruction from unconstrained photo collections}. In \bibinfo{booktitle}{\emph{Proceedings of the IEEE conference on computer vision and pattern recognition}}. \bibinfo{pages}{4197--4206}.
\newblock


\bibitem[Sailer et~al\mbox{.}(2017)]%
        {sailer2017gamification}
\bibfield{author}{\bibinfo{person}{Michael Sailer}, \bibinfo{person}{Jan~Ulrich Hense}, \bibinfo{person}{Sarah~Katharina Mayr}, {and} \bibinfo{person}{Heinz Mandl}.} \bibinfo{year}{2017}\natexlab{}.
\newblock \showarticletitle{How gamification motivates: An experimental study of the effects of specific game design elements on psychological need satisfaction}.
\newblock \bibinfo{journal}{\emph{Computers in human behavior}}  \bibinfo{volume}{69} (\bibinfo{year}{2017}), \bibinfo{pages}{371--380}.
\newblock


\bibitem[Saito et~al\mbox{.}(2019)]%
        {saito2019pifu}
\bibfield{author}{\bibinfo{person}{Shunsuke Saito}, \bibinfo{person}{Zeng Huang}, \bibinfo{person}{Ryota Natsume}, \bibinfo{person}{Shigeo Morishima}, \bibinfo{person}{Angjoo Kanazawa}, {and} \bibinfo{person}{Hao Li}.} \bibinfo{year}{2019}\natexlab{}.
\newblock \showarticletitle{Pifu: Pixel-aligned implicit function for high-resolution clothed human digitization}. In \bibinfo{booktitle}{\emph{Proceedings of the IEEE/CVF international conference on computer vision}}. \bibinfo{pages}{2304--2314}.
\newblock


\bibitem[Sanyal et~al\mbox{.}(2019)]%
        {sanyal2019learning}
\bibfield{author}{\bibinfo{person}{Soubhik Sanyal}, \bibinfo{person}{Timo Bolkart}, \bibinfo{person}{Haiwen Feng}, {and} \bibinfo{person}{Michael~J Black}.} \bibinfo{year}{2019}\natexlab{}.
\newblock \showarticletitle{Learning to regress 3D face shape and expression from an image without 3D supervision}. In \bibinfo{booktitle}{\emph{Proceedings of the IEEE/CVF Conference on Computer Vision and Pattern Recognition}}. \bibinfo{pages}{7763--7772}.
\newblock


\bibitem[Sengupta et~al\mbox{.}(2021)]%
        {sengupta2021light}
\bibfield{author}{\bibinfo{person}{Soumyadip Sengupta}, \bibinfo{person}{Brian Curless}, \bibinfo{person}{Ira Kemelmacher-Shlizerman}, {and} \bibinfo{person}{Steven~M Seitz}.} \bibinfo{year}{2021}\natexlab{}.
\newblock \showarticletitle{A light stage on every desk}. In \bibinfo{booktitle}{\emph{Proceedings of the IEEE/CVF International Conference on Computer Vision}}. \bibinfo{pages}{2420--2429}.
\newblock


\bibitem[Sengupta et~al\mbox{.}(2018)]%
        {sengupta2018sfsnet}
\bibfield{author}{\bibinfo{person}{Soumyadip Sengupta}, \bibinfo{person}{Angjoo Kanazawa}, \bibinfo{person}{Carlos~D. Castillo}, {and} \bibinfo{person}{David~W. Jacobs}.} \bibinfo{year}{2018}\natexlab{}.
\newblock \showarticletitle{SfSNet: Learning Shape, Refectance and Illuminance of Faces in the Wild}. In \bibinfo{booktitle}{\emph{Computer Vision and Pattern Regognition (CVPR)}}.
\newblock


\bibitem[Sethian(1999)]%
        {sethian1999fast}
\bibfield{author}{\bibinfo{person}{James~A Sethian}.} \bibinfo{year}{1999}\natexlab{}.
\newblock \showarticletitle{Fast marching methods}.
\newblock \bibinfo{journal}{\emph{SIAM review}} \bibinfo{volume}{41}, \bibinfo{number}{2} (\bibinfo{year}{1999}), \bibinfo{pages}{199--235}.
\newblock


\bibitem[Sevastopolsky et~al\mbox{.}(2020)]%
        {sevastopolsky2020relightable}
\bibfield{author}{\bibinfo{person}{Artem Sevastopolsky}, \bibinfo{person}{Savva Ignatiev}, \bibinfo{person}{Gonzalo Ferrer}, \bibinfo{person}{Evgeny Burnaev}, {and} \bibinfo{person}{Victor Lempitsky}.} \bibinfo{year}{2020}\natexlab{}.
\newblock \showarticletitle{Relightable 3d head portraits from a smartphone video}.
\newblock \bibinfo{journal}{\emph{arXiv preprint arXiv:2012.09963}} (\bibinfo{year}{2020}).
\newblock


\bibitem[Shi et~al\mbox{.}(2014)]%
        {shi2014automatic}
\bibfield{author}{\bibinfo{person}{Fuhao Shi}, \bibinfo{person}{Hsiang-Tao Wu}, \bibinfo{person}{Xin Tong}, {and} \bibinfo{person}{Jinxiang Chai}.} \bibinfo{year}{2014}\natexlab{}.
\newblock \showarticletitle{Automatic acquisition of high-fidelity facial performances using monocular videos}.
\newblock \bibinfo{journal}{\emph{ACM Transactions on Graphics (TOG)}} \bibinfo{volume}{33}, \bibinfo{number}{6} (\bibinfo{year}{2014}), \bibinfo{pages}{1--13}.
\newblock


\bibitem[Shi et~al\mbox{.}(2020)]%
        {shi2020fast}
\bibfield{author}{\bibinfo{person}{Tianyang Shi}, \bibinfo{person}{Zhengxia Zuo}, \bibinfo{person}{Yi Yuan}, {and} \bibinfo{person}{Changjie Fan}.} \bibinfo{year}{2020}\natexlab{}.
\newblock \showarticletitle{Fast and robust face-to-parameter translation for game character auto-creation}. In \bibinfo{booktitle}{\emph{Proceedings of the AAAI Conference on Artificial Intelligence}}, Vol.~\bibinfo{volume}{34}. \bibinfo{pages}{1733--1740}.
\newblock


\bibitem[Slossberg et~al\mbox{.}(2022)]%
        {slossberg2022unsupervised}
\bibfield{author}{\bibinfo{person}{Ron Slossberg}, \bibinfo{person}{Ibrahim Jubran}, {and} \bibinfo{person}{Ron Kimmel}.} \bibinfo{year}{2022}\natexlab{}.
\newblock \showarticletitle{Unsupervised High-Fidelity Facial Texture Generation and Reconstruction}. In \bibinfo{booktitle}{\emph{Computer Vision -- ECCV 2022}}, \bibfield{editor}{\bibinfo{person}{Shai Avidan}, \bibinfo{person}{Gabriel Brostow}, \bibinfo{person}{Moustapha Ciss{\'e}}, \bibinfo{person}{Giovanni~Maria Farinella}, {and} \bibinfo{person}{Tal Hassner}} (Eds.). \bibinfo{publisher}{Springer Nature Switzerland}, \bibinfo{address}{Cham}, \bibinfo{pages}{212--229}.
\newblock


\bibitem[Smith et~al\mbox{.}(2020)]%
        {smith2020morphable}
\bibfield{author}{\bibinfo{person}{William~AP Smith}, \bibinfo{person}{Alassane Seck}, \bibinfo{person}{Hannah Dee}, \bibinfo{person}{Bernard Tiddeman}, \bibinfo{person}{Joshua~B Tenenbaum}, {and} \bibinfo{person}{Bernhard Egger}.} \bibinfo{year}{2020}\natexlab{}.
\newblock \showarticletitle{A morphable face albedo model}. In \bibinfo{booktitle}{\emph{Proceedings of the IEEE/CVF Conference on Computer Vision and Pattern Recognition}}. \bibinfo{pages}{5011--5020}.
\newblock


\bibitem[Tewari et~al\mbox{.}(2019)]%
        {tewari2019fml}
\bibfield{author}{\bibinfo{person}{Ayush Tewari}, \bibinfo{person}{Florian Bernard}, \bibinfo{person}{Pablo Garrido}, \bibinfo{person}{Gaurav Bharaj}, \bibinfo{person}{Mohamed Elgharib}, \bibinfo{person}{Hans-Peter Seidel}, \bibinfo{person}{Patrick P{\'e}rez}, \bibinfo{person}{Michael Z{\"o}llhofer}, {and} \bibinfo{person}{Christian Theobalt}.} \bibinfo{year}{2019}\natexlab{}.
\newblock \showarticletitle{Fml: Face model learning from videos}. In \bibinfo{booktitle}{\emph{Proceedings of the IEEE Conference on Computer Vision and Pattern Recognition}}. \bibinfo{pages}{10812--10822}.
\newblock


\bibitem[Tewari et~al\mbox{.}(2022)]%
        {tewari2022advances}
\bibfield{author}{\bibinfo{person}{A. Tewari}, \bibinfo{person}{J. Thies}, \bibinfo{person}{B. Mildenhall}, \bibinfo{person}{P. Srinivasan}, \bibinfo{person}{E. Tretschk}, \bibinfo{person}{W. Yifan}, \bibinfo{person}{C. Lassner}, \bibinfo{person}{V. Sitzmann}, \bibinfo{person}{R. Martin-Brualla}, \bibinfo{person}{S. Lombardi}, \bibinfo{person}{T. Simon}, \bibinfo{person}{C. Theobalt}, \bibinfo{person}{M. Nießner}, \bibinfo{person}{J.~T. Barron}, \bibinfo{person}{G. Wetzstein}, \bibinfo{person}{M. Zollhöfer}, {and} \bibinfo{person}{V. Golyanik}.} \bibinfo{year}{2022}\natexlab{}.
\newblock \showarticletitle{Advances in {Neural} {Rendering}}.
\newblock \bibinfo{journal}{\emph{Computer Graphics Forum}} \bibinfo{volume}{41}, \bibinfo{number}{2} (\bibinfo{year}{2022}), \bibinfo{pages}{703--735}.
\newblock
\newblock
\shownote{\_eprint: https://onlinelibrary.wiley.com/doi/pdf/10.1111/cgf.14507}.


\bibitem[Tewari et~al\mbox{.}(2018)]%
        {tewari2018high}
\bibfield{author}{\bibinfo{person}{Ayush Tewari}, \bibinfo{person}{Michael Zollhoefer}, \bibinfo{person}{Florian Bernard}, \bibinfo{person}{Pablo Garrido}, \bibinfo{person}{Hyeongwoo Kim}, \bibinfo{person}{Patrick Perez}, {and} \bibinfo{person}{Christian Theobalt}.} \bibinfo{year}{2018}\natexlab{}.
\newblock \showarticletitle{High-fidelity monocular face reconstruction based on an unsupervised model-based face autoencoder}.
\newblock \bibinfo{journal}{\emph{IEEE transactions on pattern analysis and machine intelligence}} \bibinfo{volume}{42}, \bibinfo{number}{2} (\bibinfo{year}{2018}), \bibinfo{pages}{357--370}.
\newblock


\bibitem[Tewari et~al\mbox{.}(2017)]%
        {tewari2017mofa}
\bibfield{author}{\bibinfo{person}{Ayush Tewari}, \bibinfo{person}{Michael Zollhöfer}, \bibinfo{person}{Hyeongwoo Kim}, \bibinfo{person}{Pablo Garrido}, \bibinfo{person}{Florian Bernard}, \bibinfo{person}{Patrick Pérez}, {and} \bibinfo{person}{Christian Theobalt}.} \bibinfo{year}{2017}\natexlab{}.
\newblock \showarticletitle{MoFA: Model-Based Deep Convolutional Face Autoencoder for Unsupervised Monocular Reconstruction}. In \bibinfo{booktitle}{\emph{2017 IEEE International Conference on Computer Vision (ICCV)}}. \bibinfo{pages}{3735--3744}.
\newblock


\bibitem[{The Blender Foundation}(2023)]%
        {blender}
\bibfield{author}{\bibinfo{person}{{The Blender Foundation}}.} \bibinfo{year}{2023}\natexlab{}.
\newblock \bibinfo{booktitle}{\emph{Blender}}.
\newblock
\urldef\tempurl%
\url{https://www.blender.org/}
\showURL{%
\tempurl}


\bibitem[Tran et~al\mbox{.}(2019)]%
        {tran2019towards}
\bibfield{author}{\bibinfo{person}{Luan Tran}, \bibinfo{person}{Feng Liu}, {and} \bibinfo{person}{Xiaoming Liu}.} \bibinfo{year}{2019}\natexlab{}.
\newblock \showarticletitle{Towards High-Fidelity Nonlinear 3D Face Morphable Model}. In \bibinfo{booktitle}{\emph{Proceedings of the IEEE/CVF Conference on Computer Vision and Pattern Recognition (CVPR)}}.
\newblock


\bibitem[Tran and Liu(2018)]%
        {tran2018nonlinear}
\bibfield{author}{\bibinfo{person}{L. Tran} {and} \bibinfo{person}{X. Liu}.} \bibinfo{year}{2018}\natexlab{}.
\newblock \showarticletitle{Nonlinear 3D Face Morphable Model}. In \bibinfo{booktitle}{\emph{2018 IEEE/CVF Conference on Computer Vision and Pattern Recognition (CVPR)}}. \bibinfo{publisher}{IEEE Computer Society}, \bibinfo{address}{Los Alamitos, CA, USA}, \bibinfo{pages}{7346--7355}.
\newblock


\bibitem[Tretschk et~al\mbox{.}(2023)]%
        {tretschk2023state}
\bibfield{author}{\bibinfo{person}{Edith Tretschk}, \bibinfo{person}{Navami Kairanda}, \bibinfo{person}{Mallikarjun BR}, \bibinfo{person}{Rishabh Dabral}, \bibinfo{person}{Adam Kortylewski}, \bibinfo{person}{Bernhard Egger}, \bibinfo{person}{Marc Habermann}, \bibinfo{person}{Pascal Fua}, \bibinfo{person}{Christian Theobalt}, {and} \bibinfo{person}{Vladislav Golyanik}.} \bibinfo{year}{2023}\natexlab{}.
\newblock \showarticletitle{State of the Art in Dense Monocular Non-Rigid 3D Reconstruction}. In \bibinfo{booktitle}{\emph{Computer Graphics Forum}}, Vol.~\bibinfo{volume}{42}. Wiley Online Library, \bibinfo{pages}{485--520}.
\newblock


\bibitem[Wang et~al\mbox{.}(2022)]%
        {wang2022faceverse}
\bibfield{author}{\bibinfo{person}{Lizhen Wang}, \bibinfo{person}{Zhiyuan Chen}, \bibinfo{person}{Tao Yu}, \bibinfo{person}{Chenguang Ma}, \bibinfo{person}{Liang Li}, {and} \bibinfo{person}{Yebin Liu}.} \bibinfo{year}{2022}\natexlab{}.
\newblock \showarticletitle{Faceverse: a fine-grained and detail-controllable 3d face morphable model from a hybrid dataset}. In \bibinfo{booktitle}{\emph{Proceedings of the IEEE/CVF Conference on Computer Vision and Pattern Recognition}}. \bibinfo{pages}{20333--20342}.
\newblock


\bibitem[Wang et~al\mbox{.}(2023)]%
        {wang2023sunstage}
\bibfield{author}{\bibinfo{person}{Yifan Wang}, \bibinfo{person}{Aleksander Holynski}, \bibinfo{person}{Xiuming Zhang}, {and} \bibinfo{person}{Xuaner Zhang}.} \bibinfo{year}{2023}\natexlab{}.
\newblock \showarticletitle{Sunstage: Portrait reconstruction and relighting using the sun as a light stage}. In \bibinfo{booktitle}{\emph{Proceedings of the IEEE/CVF Conference on Computer Vision and Pattern Recognition}}. \bibinfo{pages}{20792--20802}.
\newblock


\bibitem[Wang et~al\mbox{.}(2021)]%
        {wang2021learning}
\bibfield{author}{\bibinfo{person}{Ziyan Wang}, \bibinfo{person}{Timur Bagautdinov}, \bibinfo{person}{Stephen Lombardi}, \bibinfo{person}{Tomas Simon}, \bibinfo{person}{Jason Saragih}, \bibinfo{person}{Jessica Hodgins}, {and} \bibinfo{person}{Michael Zollhofer}.} \bibinfo{year}{2021}\natexlab{}.
\newblock \showarticletitle{Learning Compositional Radiance Fields of Dynamic Human Heads}. In \bibinfo{booktitle}{\emph{Proceedings of the IEEE/CVF Conference on Computer Vision and Pattern Recognition (CVPR)}}. \bibinfo{pages}{5704--5713}.
\newblock


\bibitem[Ward(2004)]%
        {ward2004game}
\bibfield{author}{\bibinfo{person}{Antony Ward}.} \bibinfo{year}{2004}\natexlab{}.
\newblock \bibinfo{booktitle}{\emph{Game character development with maya}}.
\newblock \bibinfo{publisher}{New Riders}.
\newblock


\bibitem[Wu et~al\mbox{.}(2019)]%
        {wu2019mvfnet}
\bibfield{author}{\bibinfo{person}{F. Wu}, \bibinfo{person}{L. Bao}, \bibinfo{person}{Y. Chen}, \bibinfo{person}{Y. Ling}, \bibinfo{person}{Y. Song}, \bibinfo{person}{S. Li}, \bibinfo{person}{K. Ngan}, {and} \bibinfo{person}{W. Liu}.} \bibinfo{year}{2019}\natexlab{}.
\newblock \showarticletitle{MVF-Net: Multi-View 3D Face Morphable Model Regression}. In \bibinfo{booktitle}{\emph{2019 IEEE/CVF Conference on Computer Vision and Pattern Recognition (CVPR)}}. \bibinfo{publisher}{IEEE Computer Society}, \bibinfo{address}{Los Alamitos, CA, USA}, \bibinfo{pages}{959--968}.
\newblock


\bibitem[Wu et~al\mbox{.}(2023a)]%
        {wu2023deep}
\bibfield{author}{\bibinfo{person}{Jane Wu}, \bibinfo{person}{Michael Bao}, \bibinfo{person}{Xinwei Yao}, {and} \bibinfo{person}{Ronald Fedkiw}.} \bibinfo{year}{2023}\natexlab{a}.
\newblock \showarticletitle{Deep Energies for Estimating Three-Dimensional Facial Pose and Expression}.
\newblock \bibinfo{journal}{\emph{Communications on Applied Mathematics and Computation}} (\bibinfo{year}{2023}), \bibinfo{pages}{1--25}.
\newblock


\bibitem[Wu et~al\mbox{.}(2023b)]%
        {wu2023lpff}
\bibfield{author}{\bibinfo{person}{Yiqian Wu}, \bibinfo{person}{Jing Zhang}, \bibinfo{person}{Hongbo Fu}, {and} \bibinfo{person}{Xiaogang Jin}.} \bibinfo{year}{2023}\natexlab{b}.
\newblock \showarticletitle{LPFF: A Portrait Dataset for Face Generators Across Large Poses}.
\newblock \bibinfo{journal}{\emph{arXiv preprint arXiv:2303.14407}} (\bibinfo{year}{2023}).
\newblock


\bibitem[Yariv et~al\mbox{.}(2020)]%
        {yariv2020idr}
\bibfield{author}{\bibinfo{person}{Lior Yariv}, \bibinfo{person}{Yoni Kasten}, \bibinfo{person}{Dror Moran}, \bibinfo{person}{Meirav Galun}, \bibinfo{person}{Matan Atzmon}, \bibinfo{person}{Basri Ronen}, {and} \bibinfo{person}{Yaron Lipman}.} \bibinfo{year}{2020}\natexlab{}.
\newblock \showarticletitle{Multiview Neural Surface Reconstruction by Disentangling Geometry and Appearance}. In \bibinfo{booktitle}{\emph{Advances in Neural Information Processing Systems}}, \bibfield{editor}{\bibinfo{person}{H.~Larochelle}, \bibinfo{person}{M.~Ranzato}, \bibinfo{person}{R.~Hadsell}, \bibinfo{person}{M.F. Balcan}, {and} \bibinfo{person}{H.~Lin}} (Eds.), Vol.~\bibinfo{volume}{33}. \bibinfo{publisher}{Curran Associates, Inc.}, \bibinfo{pages}{2492--2502}.
\newblock


\bibitem[Yu et~al\mbox{.}(2018)]%
        {yu2018bisenet}
\bibfield{author}{\bibinfo{person}{Changqian Yu}, \bibinfo{person}{Jingbo Wang}, \bibinfo{person}{Chao Peng}, \bibinfo{person}{Changxin Gao}, \bibinfo{person}{Gang Yu}, {and} \bibinfo{person}{Nong Sang}.} \bibinfo{year}{2018}\natexlab{}.
\newblock \showarticletitle{BiSeNet: Bilateral Segmentation Network for Real-time Semantic Segmentation}. In \bibinfo{booktitle}{\emph{Proceedings of the European Conference on Computer Vision (ECCV)}}.
\newblock


\bibitem[Zeng et~al\mbox{.}(2019)]%
        {zeng2019df2net}
\bibfield{author}{\bibinfo{person}{Xiaoxing Zeng}, \bibinfo{person}{Xiaojiang Peng}, {and} \bibinfo{person}{Yu Qiao}.} \bibinfo{year}{2019}\natexlab{}.
\newblock \showarticletitle{DF2Net: A Dense-Fine-Finer Network for Detailed 3D Face Reconstruction}. In \bibinfo{booktitle}{\emph{Proceedings of the IEEE International Conference on Computer Vision}}. \bibinfo{pages}{2315--2324}.
\newblock


\bibitem[Zhang et~al\mbox{.}(2022b)]%
        {zhang2022fdnerf}
\bibfield{author}{\bibinfo{person}{Jingbo Zhang}, \bibinfo{person}{Xiaoyu Li}, \bibinfo{person}{Ziyu Wan}, \bibinfo{person}{Can Wang}, {and} \bibinfo{person}{Jing Liao}.} \bibinfo{year}{2022}\natexlab{b}.
\newblock \showarticletitle{Fdnerf: Few-shot dynamic neural radiance fields for face reconstruction and expression editing}. In \bibinfo{booktitle}{\emph{SIGGRAPH Asia 2022 Conference Papers}}. \bibinfo{pages}{1--9}.
\newblock


\bibitem[Zhang et~al\mbox{.}(2004)]%
        {zhang-siggraph2004-stfaces}
\bibfield{author}{\bibinfo{person}{Li Zhang}, \bibinfo{person}{Noah Snavely}, \bibinfo{person}{Brian Curless}, {and} \bibinfo{person}{Steven~M. Seitz}.} \bibinfo{year}{2004}\natexlab{}.
\newblock \showarticletitle{Spacetime Faces: High-Resolution Capture for Modeling and Animation}. In \bibinfo{booktitle}{\emph{ACM Annual Conference on Computer Graphics}} (Los Angeles, CA). \bibinfo{pages}{548--558}.
\newblock


\bibitem[Zhang et~al\mbox{.}(2022c)]%
        {zhang2022video}
\bibfield{author}{\bibinfo{person}{Longwen Zhang}, \bibinfo{person}{Chuxiao Zeng}, \bibinfo{person}{Qixuan Zhang}, \bibinfo{person}{Hongyang Lin}, \bibinfo{person}{Ruixiang Cao}, \bibinfo{person}{Wei Yang}, \bibinfo{person}{Lan Xu}, {and} \bibinfo{person}{Jingyi Yu}.} \bibinfo{year}{2022}\natexlab{c}.
\newblock \showarticletitle{Video-driven neural physically-based facial asset for production}.
\newblock \bibinfo{journal}{\emph{ACM Transactions on Graphics (TOG)}} \bibinfo{volume}{41}, \bibinfo{number}{6} (\bibinfo{year}{2022}), \bibinfo{pages}{1--16}.
\newblock


\bibitem[Zhang et~al\mbox{.}(2022a)]%
        {zhang2022physically}
\bibfield{author}{\bibinfo{person}{Zhenyu Zhang}, \bibinfo{person}{Yanhao Ge}, \bibinfo{person}{Ying Tai}, \bibinfo{person}{Weijian Cao}, \bibinfo{person}{Renwang Chen}, \bibinfo{person}{Kunlin Liu}, \bibinfo{person}{Hao Tang}, \bibinfo{person}{Xiaoming Huang}, \bibinfo{person}{Chengjie Wang}, \bibinfo{person}{Zhifeng Xie}, {et~al\mbox{.}}} \bibinfo{year}{2022}\natexlab{a}.
\newblock \showarticletitle{Physically-guided Disentangled Implicit Rendering for 3D Face Modeling}. In \bibinfo{booktitle}{\emph{Proceedings of the IEEE/CVF Conference on Computer Vision and Pattern Recognition}}. \bibinfo{pages}{20353--20363}.
\newblock


\bibitem[Zheng et~al\mbox{.}(2023b)]%
        {zheng2023neuface}
\bibfield{author}{\bibinfo{person}{Mingwu Zheng}, \bibinfo{person}{Haiyu Zhang}, \bibinfo{person}{Hongyu Yang}, {and} \bibinfo{person}{Di Huang}.} \bibinfo{year}{2023}\natexlab{b}.
\newblock \showarticletitle{NeuFace: Realistic 3D Neural Face Rendering from Multi-view Images}. In \bibinfo{booktitle}{\emph{Proceedings of the IEEE/CVF Conference on Computer Vision and Pattern Recognition}}. \bibinfo{pages}{16868--16877}.
\newblock


\bibitem[Zheng et~al\mbox{.}(2023a)]%
        {zheng2023pointavatar}
\bibfield{author}{\bibinfo{person}{Yufeng Zheng}, \bibinfo{person}{Wang Yifan}, \bibinfo{person}{Gordon Wetzstein}, \bibinfo{person}{Michael~J Black}, {and} \bibinfo{person}{Otmar Hilliges}.} \bibinfo{year}{2023}\natexlab{a}.
\newblock \showarticletitle{Pointavatar: Deformable point-based head avatars from videos}. In \bibinfo{booktitle}{\emph{Proceedings of the IEEE/CVF Conference on Computer Vision and Pattern Recognition}}. \bibinfo{pages}{21057--21067}.
\newblock


\bibitem[Zhu et~al\mbox{.}(2016)]%
        {zhu2016face}
\bibfield{author}{\bibinfo{person}{X. Zhu}, \bibinfo{person}{Z. Lei}, \bibinfo{person}{X. Liu}, \bibinfo{person}{H. Shi}, {and} \bibinfo{person}{S.~Z. Li}.} \bibinfo{year}{2016}\natexlab{}.
\newblock \showarticletitle{Face Alignment Across Large Poses: A 3D Solution}. In \bibinfo{booktitle}{\emph{2016 IEEE Conference on Computer Vision and Pattern Recognition (CVPR)}}. \bibinfo{publisher}{IEEE Computer Society}, \bibinfo{address}{Los Alamitos, CA, USA}, \bibinfo{pages}{146--155}.
\newblock
\showISSN{1063-6919}


\bibitem[Zollh{\"o}fer et~al\mbox{.}(2018)]%
        {zollhofer2018state}
\bibfield{author}{\bibinfo{person}{Michael Zollh{\"o}fer}, \bibinfo{person}{Justus Thies}, \bibinfo{person}{Pablo Garrido}, \bibinfo{person}{Derek Bradley}, \bibinfo{person}{Thabo Beeler}, \bibinfo{person}{Patrick P{\'e}rez}, \bibinfo{person}{Marc Stamminger}, \bibinfo{person}{Matthias Nie{\ss}ner}, {and} \bibinfo{person}{Christian Theobalt}.} \bibinfo{year}{2018}\natexlab{}.
\newblock \showarticletitle{State of the art on monocular 3D face reconstruction, tracking, and applications}. In \bibinfo{booktitle}{\emph{Computer Graphics Forum}}, Vol.~\bibinfo{volume}{37}. Wiley Online Library, \bibinfo{pages}{523--550}.
\newblock


\end{thebibliography}

\end{document}